\documentclass[a4paper,10pt]{article}
\usepackage{ucs}
\usepackage[utf8]{inputenc}
\usepackage{amsmath}
\usepackage{amsfonts}
\usepackage{amssymb}
\usepackage{amsthm}
\usepackage{rotating}
\usepackage{xcolor}
\usepackage[english]{babel}
\usepackage[T1]{fontenc}
\usepackage{graphicx}
\usepackage{hyperref}
\usepackage{cuted}
\usepackage{verbatim}
\usepackage{subfigure}
\usepackage{multirow}
\usepackage{bbold}
\usepackage{slashed}
\usepackage{indentfirst}
\usepackage{tcolorbox}
\usepackage{placeins}
\usepackage[a4paper, total={7in, 10in}]{geometry}
\usepackage{setspace}

\newcounter{eq}

\newcommand{\bpsi}{\bar{\psi}}

\begin{document}

\large

\title{\bf 
Constituent quark axial current couplings to 
light vector mesons in the vacuum and with a weak magnetic field
}

\author{ Fabio L. Braghin 
\\
{\normalsize Instituto de F\'\i sica, Federal University of Goias}
\\
{\normalsize Av. Esperan\c ca, s/n,
 74690-900, Goi\^ania, GO, Brazil }}
\date{}

\maketitle

\begin{abstract}
Unusual constituent quark axial current couplings 
to light
vector mesons, $\rho$ and $\omega$,
are derived
in the vacuum and under   weak magnetic field
by considering a quark-antiquark interaction mediated by a 
non perturbative gluon exchange.
Similarly,
light axial mesons are found to couple anomalously with the
constituent  quark vector current.
These interactions  
are  of the type of  the Wess-Zumino-Witten terms,
being  strongly anisotropic 
and dependent
on the vector (or axial) meson polarization.
They also provide axial (vector)  form factors for the vector (axial) mesons
and  are quite small, 
 suppressed nearly 
by $1/{M^*}^2$ with respect to the 
vector mesons minimal coupling to the quark vector current.
 Some three leg meson vertices are also presented:
  $\pi-\rho-A_1$ and $V_1V_2 A$  (where $V_1,V_2$
are  vector mesons and $A$  an axial meson).
A vector and axial-vector mesons mixing is identified at non zero magnetic field 
which however can contribute only in the presence of a third particle or in a medium. 
Numerical results are presented for different 
   effective   gluon  propagators.
\end{abstract}

\section{ Introduction}
\label{intro}

{Pions, neutrinos  and eventually the light axial mesons
can be
considered  to probe the   axial content of the nucleon 
or their corresponding constituent quarks
and  of  other hadrons.}
Nucleon's axial form factor 
provides a measure of its  spin content 
and the interplay of strong and 
weak charges and interactions.
It has been
 extensively investigated theoretically and experimentally, for few works
\cite{exp-r2A,gaillard-savage,choi-etal-93,bardin-etal-1981,andreev-etal-2007,beise,drechsel-walcher,exp-bernard+E+meissner,maris-craig-tandy,eichmann-fischer,ball-chiu,segovia-etal}.
 The decomposition of the nucleon, or correspondingly constituent quark, axial vertex
involves eight tensor structures
\cite{eichmann-fischer,ball-chiu}.
The physical processes and particles that can probe them
can be
   attached to each of these structures. 
 Mesons form factors, however, are far more difficult to be measured.
Some calculations for spacelike or timelike rho-meson form factors can be found, for example, in 
\cite{hlroberts-etal,vecmesff1,vecmesff2,PRD-2018a}.At the same time,
the understanding of hadron dynamics depend on different
types of  interactions.
 For instance,
it becomes important to understand how these mesons not only couple
to   the nucleon but also
how  interactions among mesons can manifest in experimental situations.
Besides that, the way these couplings might be related to each other
might be of relevance for establishing connections to 
QCD, eventually as manifestation of fundamental symmetries.
It is expected that eventual breakdowns of fundamental symmetries should be
observed with higher precision experiments.

Low energies (Strong) couplings  of  the 
constituent quark (or nucleon)
axial 
 current involve either
the pion or the light axial mesons.
Among these axial mesons the $A_1$ and the $f_1$ 
are usually considered to be chiral partners to the $\rho$ and $\omega$
\cite{eLSM,PDG}.
There  are   however  difficulties to determine
axial mesons properties, in particular because they 
are highly unstable.  Therefore there are uncertanties about
their structures \cite{kanchen-etal,du-zhao}.
  In
Ref. \cite{du-zhao}
several couplings and decays  of the $A_1$ were analyzed with possible mixing angles.
To understand further these mesons it becomes important to 
take into account how they interact with other hadrons before decaying.
Light axial mesons couplings and decays, including its decay to the pion and a vector meson,
have also been found to be relevant to describe the $\tau$-decay
by the axial current \cite{taudecay1,taudecay2,wagner-leupold,Mikhasenko-etal}.
It has also been considered 
for the $\mu$ (g-2) problem \cite{g2-problem}
wherein the VVA correlators are important.
Three-leg mesons vertices, many time representing decays, can also be
searched experimentally and this is planned in different facilities 
as for example at FAIR/GSI \cite{f1285prod}.
Eventually, associated information from finite energy density medium 
should be compatible with their dynamics in the vacuum.
As an example, the $\rho-\omega-A_1$ and $\rho-\pi-A_1$ couplings give
 rise to the 
in medium mixing $\rho-A_1$
\cite{sasaki-plb,rho-a1-1,wagner-leupold,harada-etal}.
Eventually this mixing contributes  for the dilepton spectrum in
hot and/or dense matter  \cite{sasaki-etal1,sasaki-plb}.
The role of vector and axial mesons at finite energy density
up to  the Chiral phase transition
has been investigated in different approaches  \cite{tripolt-etal,kovacs-etal}. 
Besides that, further meson interactions might lead to modifications in the 
nuclear potential \cite{nuclear}.
From the strict theoretical point of view, three leg couplings, such as  the $\rho-\pi-A_1$
or $\rho-\omega-\pi$,
are suppressed in the $1/N_c$ counting.
Nevertheless, they have been  found to arise in different approaches
such as in  the massive Yang Mills gauge theory,
or even 
as Wess-Zumino-Witten (WZW) terms \cite{adler-bardeen,witten-1983},
in 
 \cite{gomm-etal,vecmes,hohler-rapp,ER-1986,sigmamodel,wagner-leupold}.
Facilities currently working and being built are expected to probe several aspects
of vector and axial mesons and their interactions.
High density investigations on the restoration of chiral symmetry 
and meson spectroscopy 
are planned in FAIR/GSI,  NICA and J-PARC.
Vector meson photoproduction and production mostly in ultraperipheral collisions
have been investigated in HERMES, JLAB,  FAIR/GSI   and SLAC
and currently in LHC besides plans for LHC and EIC.
Polarized vector mesons  are currently produced and investigated
in  JLAB for example in
 \cite{vecmes-photo,roy-thesis}.
Besides that axial meson production in central collisions has also been envisaged
\cite{lebiedowicz-etal}.

The constituent quark model (CQM) 
 remains one of the   leading pictures for global properties of 
hadron structure around which other contributions can  be evaluated,
for example in
\cite{ECQM-miller-etal,diquarks-craig,5quarks}.
Several nucleon  properties are described 
directly, and mostly,  in terms of constituent quarks that carry 
masses and charges of hadrons
and they are directly responsible for nucleon's couplings to mesons and 
other particles. 
This association might   somewhat manifest in 
baryons form factors.   A more precise relation between this 
two levels of description can be helpful to the detailed understanding
of the hadrons structure and dynamics.
The lack of  understanding of  the confinement mechanism does not
prevent a good theoretical description of hadron properties and interactions.
It  becomes then interesting 
to understand further  predictions of these CQM.
Global Color  (GCM) \cite{PRC1,tandy} and 
Nambu-Jona-Lasinio (NJL)
\cite{klevansky,weise-vogl} 
 type models
go along the CQM  description in terms of constituent  quarks.
They have been shown to be
 suitable for the description of many aspects of light meson  spectrum and properties.
{ In these models, Dynamical Chiral Symmetry Breakdown (DChSB)
leads to the formation of  chiral condensates, 
directly  associated as sea quarks states
  \cite{BRST-2010}.
DChSB endows 
quarks  with a large mass 
providing an important link between fundamental symmetries at the QCD 
and the hadron levels.}
 In the CQM,  the  pion, as a  Goldstone boson, has 
the    axial coupling to (constituent) quarks whose coupling constant   $g_A$
 is usually given  by $g_A = 3/4$, $g_A = 5/3$ or $g_A=1$
 \cite{weise-vogl,weinberg-axial,weinberg-2010}.
  Light mesons couplings to 
constituent quarks and form factors have been derived
 dynamically and analytically \cite{PRD-2018a,PRD-2019}
along with the same spirit of the  CQM.
By starting with a leading term of QCD quark-effective action,
based in one non-perturbative gluon exchange
(GCM),
standard techniques
have been applied and they 
 will be considered in the present work.
 Background quarks, dressed by a sort of gluon cloud,
give rise to constituent  quarks.
Mesons interactions with constituent quarks arise
and provide, for example,   a pion cloud.
This is seen explicitly in the  
  Weinberg's large N$_c$
effective field theory  (EFT), that copes  constituent quark picture
with 
 the  large N$_c$ expansion 
 \cite{weinberg-2010}.
This EFT
has also  been derived with symmetry breaking corrections 
with the same analytical techniques considered in the present work
in the vacuum and under weak magnetic fields
\cite{EPJA-2016,EPJA-2018}.
Furthermore, this approach might  lead to  a straightforward extension to 
finite baryon densities or finite temperatures.
 Moreover, the approach considered directly provides
relative strength (ratios) of  different coupling constants or form factors.
These ratios lead to reasonable estimates of their relative 
importance. 
This is usually important also for  planning experiments or 
for the interpretation of results.

Besides the interest in calculating hadron couplings in the vacuum,
strong magnetic fields have been estimated to appear 
in non central relativistic heavy ions collisions (r.h.i.c.) and magnetars
and they can produce many different effects
 \cite{review-B-1,review-B-2}.
Although these magnetic fields can be as large 
as $10^{15}$ T, they are not so large if compared to an hadron mass scale
such as an effective (constituent) quark mass - or even the pion mass.
One has typically  $(eB_0) \sim 10^{-2}{M^*}^2 - 10{M^*}^2$, 
where $M^*$ is a quark effective mass typical from the CQM.
It becomes interesting to identify changes in the quark and hadron dynamics
due to the external magnetic fields. 
Under (relatively) weak magnetic fields,  pions, vector and axial mesons 
develop additional couplings to (constituent) 
quark currents
 \cite{PRD-2018b,JPG-2020a}.
These couplings may have   reduced strengths 
as compared to the couplings to   quark  vector and axial currents.
 At energies in which magnetic fields can be expected to  show up in 
non-central heavy ions collisions, vector and axial mesons 
can also be more copiously  produced.
 Therefore the investigation of the effects 
of magnetic fields in the  mesons dynamics and couplings becomes 
of further interest.

In the present work,
 light vector mesons (rho and omega)
anomalous
couplings to the constituent quark
axial current will be  derived in the vacuum and under a constant weak
magnetic field
by considering the same  framework used in 
\cite{PRD-2018a,PRD-2019,PRD-2018b,JPG-2020a,JPG-2020b}.
These anomalous couplings can be considered
as corrections to the rho/omega ($A_1/f_1$) form factors
that correspond to a very small axial (vector)  component.
As such, this small axial component should be at the origin of the 
rho-$A_1$  mixing expected to occur in the medium.
The same can be expected for a $\omega-f_1$ mixing.
They can be seen as  Wess-Zumino-Witten  type terms.
These coupling functions are  considerably smaller than the known rho 
vector meson coupling to the nucleon (constituent quark) vector current.
Besides that, anomalous
three-leg mesons couplings that might correspond or contribute for 
vector and/or axial mesons mixings will also be presented.
The work is organized as follows.
In the next section, the main steps of the approach are
briefly reminded  and  the quark  determinant will be exhibited.
A large quark effective  mass  expansion is performed and the (next leading order)
{\it anomalous}  terms will be presented.
The  next  leading terms, and form factors,
 are
presented in sections (\ref{sec:two-Q}) and (\ref{sec:Bfield}).
These couplings are shortly compared
to leading couplings of the pion,  light  vector and axial  mesons 
previously derived.
The action term for the vector meson coupling to the axial current, as a WZW type term,
is shown to be  topologically conserved.
 Numerical results are presented in the  section (\ref{sec:numerical}) for 
the low    momentum regime, $Q^2< 1$ GeV$^2$,
 by considering three 
different effective gluon propagators.
 On-shell values for some of the coupling constants are provided 
and estimations for  the axial  averaged quadratic radius (a.q.r.) of 
the vector  mesons are calculated.
In the last section there is a summary and final remarks.
 
\section{ Quark determinant and  next leading anomalous couplings }
\label{sec:two-Q}

A QCD-based model for quark-antiquark interaction mediated by 
a gluon is the GCM, that takes into account non perturbative/non-Abelian gluon effects 
by means of an effective  gluon propagator.
It is given by: 
\begin{equation} \label{Seff}  
 Z [\eta, \bar\eta] = N \int {\cal D}[\bpsi, \psi]
\exp \;  i \int_x  \left[
\bar{\psi} \left( i \slashed{D} 
- m \right) \psi 
- \frac{g^2}{2}\int_y j_{\mu}^\beta (x) 
{\tilde{R}}^{\mu \nu}_{\beta \alpha}  (x-y) j_{\nu}^{\alpha} (y) 
+ \bpsi \eta + \bar{\eta} \psi \right] ,
\end{equation}
Where the color  quark current is 
$j^{\mu}_\alpha = \bar{\psi} \lambda_\alpha  \gamma^{\mu} \psi$, 
 $\int_x$ 
stands for 
$\int d^4 x$,
$i,j,k=0,...(N_f^2-1)$ will be used  for, SU($N_f=2$),
isospin indices and
$\alpha,\beta...=1,...(N_c^2-1)$ stands 
for color in the adjoint representation.
The sums in color, flavor and Dirac indices are implicit
and $\eta, \bar{\eta}$ are the quark sources.
$ D_{\mu} = \partial_\mu  - i e Q A_{\mu}$
is the covariant quark derivative  with the minimal coupling to 
a background electromagnetic field,
with  the diagonal matrix 
$\hat{Q} =  diag(2/3, -1/3)$ 
{ 
for up and down electromagnetic charges.}
 To account for the non-Abelian   structure of the gluon sector the gluon propagator
$\tilde{R}^{\mu\nu}_{\alpha\beta}(x-y)$
must be non perturbative. As an external input for the model,
it will be required to  have enough strength to yield 
Dynamical Chiral Symmetry Breaking (DChSB)
with a given strength of the (running) quark-gluon coupling constant.
DChSB has been found in several works with different
approaches, for few examples 
\cite{SD-rainbow,kondo,cornwall}. 
It is one of the mechanisms that endows hadrons with large masses
with respect to the (measured) quark masses \cite{dcsb+traceanomaly}.
 Other terms  from QCD effective action, such as genuine 
three and four quark interactions, 
 due the non Abelian gluon structure, are not 
considered.
 In  several gauges 
the  gluon propagator $\tilde{R}^{\mu \nu}_{\alpha\beta}(k)$
 can be written in momentum space as:
$ \tilde{R}^{\mu\nu}_{\alpha\beta} (k) = \delta_{\alpha\beta} \left[
\left( g^{\mu\nu} - \frac{k^\mu k^\nu}{ k^2}
\right) 
 R_T  (k)
+  \frac{ k^\mu k^\nu}{ k^2} 
 R_L (k) \right]$,
where $R_T(k), R_L(k)$ are
transversal and longitudinal components.
Contributions from extra terms due to confinement and gauge boson dynamics
proportional to $\delta (p^2)$ 
\cite{lowdon} can be shown to be smaller or even vanishing   \cite{FLB-2021a}
for the observables calculated below.

By means of a Fierz transformation, 
Dirac and isospin structures 
can be suitable introduced such as to 
 provide the correct quantum numbers for leading low energies
quark-antiquark states
corresponding to mesons.
The background field method and the auxiliary field method
are then used to calculate the quark determinant
in the presence of DChSB  
by considering the auxiliary fields 
associated with quark-antiquark mesons states.
Besides that,  background quark currents give rise to 
constituent quark currents.
{
In particular, the auxiliary field method (AFM) was employed explicitly in 
Refs. \cite{PRD-2018a,PRD-2019,EPJA-2016,JPG-2020b} for bilocal auxiliary fields.
 These auxiliary fields can be expanded in a complete basis of 
 structureless mesons fields.
Field renormalization constants can be introduced explicitly 
with which the unit integral of the AFM can be written as:
\begin{eqnarray} \label{AF1}
 1 &=& N 
 \int 
D[V_\mu^i , \bar{A}_\mu^i, V_\mu, \bar{A}_\mu]
 e^{- \frac{i \alpha }{4 }
\int_{x,y} 
(\bar{R}^{\mu\nu})^{-1} ( 
(Z_V^{\frac{1}{2}} V^i_{\mu} - 
  g  Z_g Z_\psi     {\bar{R}_{\mu\nu}}  {j^{V,\nu}}^{i} )  
(Z_V^{\frac{1}{2}} V_i^{\mu} - 
  g  Z_g Z_\psi     {\bar{R}_{\mu\nu}}  {j_{V,\nu}}^{i} ) }
\nonumber
\\
&\times&
e^{- \frac{i \alpha }{4 }
\int_{x,y} 
(\bar{R}^{\mu\nu})^{-1} (Z_A^{\frac{1}{2}} \bar{A}^i_{\mu} - 
  g  Z_g Z_\psi     {\bar{R}_{\mu\nu}}  {j^{A,\nu}}^{i} )  
(Z_A^{\frac{1}{2}} \bar{A}_i^{\mu} - 
  g  Z_g Z_\psi     {\bar{R}_{\mu\nu}}  {j_{A,\nu}}^{i} ) 
}
\nonumber
\\
&\times&
  e^{- \frac{i \alpha }{4 }
\int_{x,y} 
(\bar{R}^{\mu\nu})^{-1} ( 
(Z_V^{\frac{1}{2}} V_{\mu} - 
  g  Z_g Z_\psi     {\bar{R}_{\mu\nu}}  {j^{V,\nu}} )  
(Z_V^{\frac{1}{2}} V^{\mu} - 
  g  Z_g Z_\psi     {\bar{R}_{\mu\nu}}  {j_{V,\nu}} ) }
\nonumber
\\
&\times&
e^{- \frac{i \alpha }{4 }
\int_{x,y} 
(\bar{R}^{\mu\nu})^{-1} (Z_A^{\frac{1}{2}} \bar{A}_{\mu} - 
  g  Z_g Z_\psi     {\bar{R}_{\mu\nu}}  {j^{A,\nu}} )  
(Z_A^{\frac{1}{2}} \bar{A}^{\mu} - 
  g  Z_g Z_\psi     {\bar{R}_{\mu\nu}}  {j_{A,\nu}} ) 
}
\nonumber
\\
&\times&
\int D[S_i] D[P_i]
 e^{- \frac{i}{2 }  
\int_{x,y}   R (x-y)  \alpha  \left[ ( Z_S^{\frac{1}{2}}S - g   Z_g Z_\psi   j^S_{(2)})^2 +
(Z_P^{\frac{1}{2}} P_i -  g    Z_g Z_\psi   j^{P}_{i,(2)} )^2 \right]}
,
\end{eqnarray}
where 
\begin{eqnarray} 
\bar{R}^{\mu\nu} &\equiv& \bar{R}^{\mu\nu}(x-y)= g^{\mu\nu} ( R_T(x-y) + R_L(x-y) )
+ 2 \frac{\partial^\mu \partial^\nu}{\partial^2 } ( R_T(x-y) - R_L(x-y) ) ,
\\
R &\equiv& R(x-y) = 3 R_T (z-y) + R_L (x-y).
\end{eqnarray}
The scalar and pseudoscalar fields  give rise to a non-linear representation 
in terms of the (Goldstone boson) 
pion field by a usual chiral rotation. 
The renormalization constants will not be carried on in the calculations  below,
being therefore omitted.
However, they allow for a systematic elimination of the 
ultraviolet divergences in the resulting couplings of the 
effective action.

}
The  Gaussian integration of the 
   quark field can now be performed in the presence of 
background quark currents and mesons fields that are 
arranged  in a chiral invariant way.
By making use of the identity
$\det A = \exp \; Tr \; \ln (A)$, 
it yields:
\begin{eqnarray} \label{Seff-det}  
S_{eff}   &=& - i  \; Tr  \; \ln \; \left\{
 i \left[ {S_0}^{-1} (x-y) 
+ \Xi_v  (x-y)
+ \Xi_s (x-y) 
+
\sum_q  a_q \Gamma_q j_q  (x,y) \right]
 \right\} ,
\end{eqnarray}
where 
$Tr$ stands for traces of all discrete internal indices 
and integration of  space-time coordinates.
The quantity
$\Xi_v (x-y)$ encodes the vector and axial mesons contributions 
and $\Xi_s (x-y)$ encodes the pion field contribution that arise
from the scalar and pseudoscalar auxiliary fields.
In the last term of Eq. (\ref{Seff-det}) there is a sum of 
background quark currents with  Dirac, flavor and color 
structures among  which 
only the  (isosinglet and isotriplet)
axial currents  will be kept:
\begin{eqnarray} \label{Rq-j}
\sum_q  a_q \Gamma_q  j_q (x,y)
&\to&
 -  \alpha g^2
 \bar{R}^{\mu\nu} (x-y) \gamma_\mu 
\left[  \sigma_i 
  \gamma_5   \bpsi  (y)
 \gamma_5 \gamma_\nu  \sigma_i \psi (x) 
+ 
 \gamma_5   \bpsi  (y)
 \gamma_5 \gamma_\nu   \psi (x)
\right],
\end{eqnarray}
where  $\sigma_i$ are the Pauli matrices,
$\bar{R}^{\mu\nu} = 2 [3 R_T (x-y) + R_L (x-y)]$
and $\alpha = 4/9$.
The remaining terms  can be written as:
\begin{eqnarray} \label{vectors-q}
 \Xi_v (x-y) 
&=& 
 -  
 \frac{\gamma^\mu }{2}
 \left[ 
 F_v  \sigma_i
 \left(     V_{\mu}^i (x)
+   \gamma_5 
 \bar{A}_{\mu}^i (x) 
+     V_{\mu} (x)
+   \gamma_5 
 \bar{A}_{\mu} (x) \right)
  \right]   \delta (x-y),
\\
 \Xi_s  (x-y) &=& F ( P_R  U + P_L  U^\dagger) \; \delta (x-y),
\end{eqnarray}
where 
 $U = e^{i \vec{\pi} \cdot \tau/2}$, 
$P_{R/L} =  (1 \pm \gamma_5)/2$ are the chirality 
 right/left hand projectors,
  $F_v$ 
and $F$
provide  the canonical field definitions
meson fields,
for both iso-triplet (rho and A$_1$) and iso-singlet ($\omega$ and $f_1$)
\cite{PDG,vec-ax1},  and for the pion.
$F_v$ will be incorporated into the vector and axial mesons fields to provide their 
canonical dimensions.
The other quark-antiquark mesons, their auxiliary fields, will be neglected.
 The gap equations for the auxiliary fields
are derived as saddle point equations
and solved as usually done for the model 
(\ref{Seff}) and NJL-type models.
The scalar field equation will be the only one with non trivial non zero solution
and the resulting quark-antiquark scalar condensate ($\bar{S}$)
becomes responsible for the increased quark mass $M^* = m + \bar{S}$.
The quark propagator  
can then be written as:
$ S_{0}^{-1}(x-y)  =  
\left(
  i \slashed{D}
-
  M^*  \right)
\delta (x-y)$.

A large quark mass expansion
\cite{PRD-2014} with a zero order derivative expansion \cite{integr-pathint}
 of the determinant (\ref{Seff-det})  will be implemented now.
It is  suitable for the 
long wavelength limit of the model and they 
have been performed in previous works of the author mentioned above.
The leading terms of the expansion with 
vector/axial currents and the light mesons
have been collected
and investigated previously by the author in the vacuum and under  weak magnetic fields.
Next leading  couplings of
 (iso-triplet $\rho$ and iso-singlet $\omega$)
light vector mesons
to constituent quark axial current also  appear 
as non-local interactions.
With
explicit momentum dependencies 
they  can be written as:
\begin{eqnarray}
\label{L-vja}
{\cal L}_{vja}
&=&
 i  \delta_{ij} \epsilon^{\sigma \rho \mu  \nu}
  F^{vja} (K,Q) 
K_\sigma {\cal F}^i_{\rho \mu} (Q) j^{A, j}_\nu  (K,K+Q)
\nonumber
\\
&+& 
 i   \epsilon^{\sigma \rho \mu  \nu}
  F^{vja} (K,Q) 
K_\sigma {\cal F}_{\rho \mu} (Q) j^{A}_\nu  (K,K+Q) ,
\end{eqnarray}
where $Q$ is the vector meson  4-momentum, $K$ is the  quark momentum.
The iso-triplet and iso-singlet 
axial currents were  defined as
$j^{A,i}_\mu (K,K+Q) = \bar{\psi} (K+Q) \gamma_\mu \gamma_5 \sigma^i \psi (K)$
and 
$j^{A}_\mu (K,K+Q) = \bar{\psi} (K+Q) \gamma_\mu \gamma_5 \psi (K)$.
The (Abelian limit of the) stress tensors were defined for the 
isotriplet and isosinglet states:
\begin{eqnarray}
{\cal F}_{\rho\mu}^i (Q) &=& Q_\rho V_\mu^i (Q) - Q_\mu V_\rho^i (Q),
\;\;\;\;\;\;\;\;
{\cal F}_{\rho\mu} (Q) = Q_\rho V_\mu (Q) - Q_\mu V_\rho (Q).
\end{eqnarray}
The non-Abelian contribution  was neglected although it 
can be incorporated and it 
leads to  three and four-vector or axial  mesons-quark vertices.

The above  form factor is
 given by:
\begin{eqnarray}
\label{Gvja}
F_{vja} (K,Q)  &=&
4 d_2 N_c  \; (\alpha g^2)  \;    \int_k \; 
((\tilde{S}_0 (k+K) \tilde{S}_0 (k+K+Q) \bar{\bar{R}} (-k) )) ,
\end{eqnarray}
where $\int_k = \int \frac{d^4 k}{(2 \pi)^4}$
in the Euclidean momentum space, $\bar{\bar{R}}(-k)= 2 R(-k)$,
$d_n = \frac{(-1)^{n+1} }{2 n}$.
The double parenthesis was used
for denoting the order of the original kernels
and it implicitly contains the momentum structure resulting from 
the trace in Dirac indices.
{This integral,  for infrared (IR) regular
 gluon propagators,  is   finite.}
The following function was  defined:
$\tilde{S}_{0} (k) =
  \frac{1}{ k^2 + {M^*}^2}$.
These  interactions (\ref{L-vja})
emerge due to the following {\it anomalous}
trace of Dirac matrices:
\begin{eqnarray} \label{trace=5Dirac}
Tr_D ( \gamma_\mu \gamma_\nu \gamma_\rho \gamma_\sigma \gamma_5 ) 
= 4 i \epsilon_{\mu\nu\rho\sigma},
\end{eqnarray}
where $ \epsilon^{\rho \mu \sigma \nu}$ is the Levi-Civita tensor.
Along this work, 
the momentum dependence of the form factors is written as 
$ F (K,Q)$ that is a shorthanded notation for 
$F (K^2, Q^2, K\cdot Q)$.
The momentum dependent form factors
in eq. (\ref{L-vja}) have dimensions of mass$^{-2}$.
Both vector mesons, rho and omega, couple 
to 
the corresponding axial current with the same strength $F_{vja}(K,Q)$.

The above interactions (\ref{L-vja})
 have chiral counterparts of the light axial mesons $A_1$ and $f_1$,
denoted by $\bar{A}_\mu^i$ and $\bar{A}_\mu$ respectively,
interacting with the vector quark currents.
These couplings are given by:
\begin{eqnarray}
\label{L-vja-A}
{\cal L}_{vja-A}
&=&
 i  \epsilon^{\sigma \rho \mu  \nu}
  F^{vja} (K,Q)  
K_\sigma {\cal G}^i_{\rho \mu} (Q) 
j^{V, i}_\nu  (K,K+Q)
\nonumber
\\
&+&
 i   \epsilon^{\sigma \rho \mu  \nu}
  F^{vja} (K,Q)  
K_\sigma {\cal G}_{\rho \mu} (Q) j^{V}_\nu  (K,K+Q) ,
\end{eqnarray}
where
$j^{V,i}_\mu (K,K+Q) = \bar{\psi} (K+Q) \gamma_\mu \sigma^i \psi (K)$
and 
$j^{V}_\mu (K,K+Q) = \bar{\psi} (K+Q) \gamma_\mu \psi (K)$.
The (isotriplet and isosinglet)   axial  mesons  (Abelian)
 stress tensors
in coordinate space are the following:
\begin{eqnarray}
{\cal G}_{\mu\nu}^i &=& 
\partial_\mu \bar{A}_\nu^i - \partial_\nu \bar{A}_\mu^i, 
\;\;\;\;\;\;
{\cal G}_{\mu\nu} =
\partial_\mu \bar{A}_\nu - \partial_\nu \bar{A}_\mu.
\end{eqnarray}
Non-Abelian terms can be incorporated 
 \cite{vecmes,vec-axial-mesons}.

The above anomalous   interactions  (\ref{L-vja}) and (\ref{L-vja-A})
are strongly anisotropic with respect to 
the  quark and vector mesons
momenta, and the vector meson polarization. 
They   arise in the same way 
the Wess-Zumino-Witten type-terms  that
were also obtained for example in the 
bosonized version of the NJL-model, sigma models and other approaches
 \cite{gomm-etal,ER-1986,PRC1,sigmamodel}.
The anti-symmetric tensor $\epsilon_{\sigma\rho\mu\nu}$ obviously prevents
the contributions of many components of quark  and vector meson 
momenta.
Moreover, it  makes important to select a 
particular vector (or axial)  meson polarization.
As a whole, the vertices $Q_\mu K_\nu F_{vja}(K,Q)$ 
 probe an  anisotropy  of the  
spatial variation of the  axial (vector)  current and charge distribution
in eq. (\ref{L-vja})  (eq. (\ref{L-vja-A})).
These couplings correspond to   a small anomalous axial (vector)
component of the vector (axial) meson structure.
As such, it may be responsible for vector-axial mesons mixings.
This type of mixing could only occur in the presence of a third particle
or in a medium to conserve momenta.

\subsection{ Comparison with other couplings of the axial current}

The leading light mesons (pion, omega, rho and axial $A_1(1260)$) interactions
 with the axial isotriplet quark current 
had been obtained by the same method used above
 \cite{EPJA-2016,PRD-2018a,PRD-2019}.
By 
 adding to the  anomalous coupling above, they 
can be written as:
\begin{eqnarray} \label{mesons-couplings}   
{\cal L}_{j_A} =
\left[ G_A (Q,K) Q_\mu \pi^i (Q) 
+ G_{\bar{A}} (Q,K) \bar{A}_\mu^i (Q)
+   i\; F_{vja} (Q,K) \epsilon_{\mu\nu\rho\sigma} K^\nu Q^\rho V^\sigma_i (Q)
\right]
 j_{A,i}^\mu (K,Q),
\end{eqnarray}
where a dimensionless pion field was written in the first term. 
The  analogous terms for the isosinglet axial current were omitted.
Other couplings with different - higher order - 
momentum dependence might arise being outside  of the scope of this work.
The axial meson coupling  to the axial current 
has been considered recently in different approaches \cite{unal-etal}.
The following form factors were used in eq. (\ref{mesons-couplings})
\begin{eqnarray}
 \label{GA-U} 
 G_A (K,Q)  = G_V (K,Q)  &=& 
4 F \;C_{A,V}  \;   
\int_k   
(( {M^*} \tilde{S}_0(k+K) \tilde{S}_0 (k+Q+K) \bar{\bar{R}}(-k) )),
\\
\label{gr1}
 G_{\bar{A}} (K,Q) 
&=& 
 \;  C_{A,V}  \;
  \int_k 
 \frac{ ( k +K) \cdot ( k + K+Q)   + {M^*}^2 }{ 
(  (k+K)^2 + {M^*}^2 ) ( (k+K+Q)^2 + {M^*  }^2 )}
 \bar{\bar{R}} (-k ) 
,
\end{eqnarray}
{\bf where $C_{A,V} = 4 N_c d_2 (\alpha g^2)$}
and $F$ is the pion decay constant.
The form factor $F_{vja} (K,Q)$, eq. (\ref{Gvja}), 
is suppressed
 by $\sim 1/M^*$ or $\sim 1/{M^*}^2$  with respect to $G_A$ and $G_{\bar{A}}$
and also with respect to the rho or omega-coupling to the quark vector current.
Besides that, it has a tensor structure 
 dependent on $K Q$
and 
it  presents an intrinsic anisotropy in momentum space
that contributes to make him smaller.
Note that  $G_{\bar{A}} (K,Q)$ (as well as the vector meson coupling)
 is dimensionless for the 
canonical normalization of axial-vector fields.

 It is also interesting to note that,
for  constant quark effective masses $M^*$,
the pion axial coupling, $G_A(K,Q)$,
 is directly proportional to the 
anomalous coupling constant or form factor  $F_{vja} (K,Q)$. 
They can  be related by:
\begin{eqnarray} \label{ratio-GA-FVJA}
\frac{ F_{vja} (K,Q) }{ G_A (K,Q) }
= \frac{ 1
}{ 4 M^* F
}.
\end{eqnarray}
Therefore the anomalous coupling constant $F_{vja}$ can be 
considered to be    suppressed with respect 
to $G_A \sim 1$.

\subsection{ Some three-leg mesons couplings
}

The meson sector of the 
quark determinant (\ref{Seff-det}) has been investigated extensively
by many groups, for example in \cite{vecmes,ER-1986}.
Next, the leading coupling terms that involve 
a vector meson and an axial meson are exhibited.
Although not all of them are non-zero in the vacuum
they are all displayed and their possible contributions discussed below.
The leading vector-axial mesons  couplings with a pion  can be written as:
\begin{eqnarray} \label{GVA}  
{\cal L}_{mix}
&=&
G_{v-a-\pi}  (K,Q) \left(
i \epsilon_{ijk}
\pi_i (Q+K) V_\mu^j (Q) A^\mu_k (K) +
\pi_i (Q+K) V_\mu^i (Q) A^\mu (K) \right)
\nonumber
\\
&+&  
G_{v-a-\pi}  (K,Q) \left(
\pi_i (Q+K) V_\mu (Q) A^\mu_i (K) 
\right) 
\nonumber
\\
&+&
G_{kv-a-\pi}  (K,Q)
i \epsilon_{ijk} \left(
\partial^\nu \pi_i (Q+K) Q_\mu V_\nu^j K A^\mu_k (K) 
+
\partial_\nu \pi_i (Q+K) V_\mu^i (Q) K^\mu A^\nu (K) \right)
\nonumber
\\
&+&
G_{kv-a-\pi}  (K,Q)
i \epsilon_{ijk}
  \pi_i (Q+K)  Q_\mu V_\nu^j (Q) K^\mu A^\nu_k (K) 
+ {\cal O}_k
\end{eqnarray}
where ${\cal O}_k$ contains further couplings with higher order momentum dependence,
 $Q,K$  are the momenta carried by the 
vector and  axial mesons, respectively.
In this equation the pion field is canonically normalized.
The
coupling functions are   given by:
\begin{eqnarray} \label{GVA-def}
 \label{GVAPI}
G_{v-a-\pi}  (K,Q) &=&
d_2 N_c 6 \int_k 
(( M^* T(K,Q)
\tilde{S}_0 (k) \tilde{S}_0 (k+K)
\tilde{S}_0 (k+K+Q) ))
,
\\
G_{kv-a-\pi} (K,Q) &=&  d_2 N_c 6 \int_k 
(( M^*
\tilde{S}_0 (k) \tilde{S}_0 (k+K)
\tilde{S}_0 (k+K+Q) )) ,
\end{eqnarray}
where $T(K,Q) \equiv (2  k\cdot (k+K) + K \cdot (K+Q) -  {M^*}^2)/2$.
 Similar couplings to the  $\rho-\pi-A_1$ 
 have been proposed in other works
within  different or similar approaches, for example in
 \cite{ecker-etal-1989,roca-etal-2004,vecmes,taudecay1,taudecay2,wagner-leupold,sigmamodel,birse}
and references therein.
The integral of the coupling function
$G_{v-a-\pi}$  is
  ultraviolet  (UV) logarithmic divergent
and  $G_{kv-a-\pi}$ is UV finite.
They have respectively   dimensions  of mass and mass$^{-1}$.

This UV divergence in eq. (\ref{GVAPI})  is the same
of the momentum dependent free
vector and axial mesons terms.
The free and self interacting light vector and axial mesons -sector
have been investigated in very similar approaches,
see for example in \cite{vecmes,tandy-maris,vec-axial-mesons}.
This divergence  is directly eliminated by the vector and axial fields
renormalization constants in the same way some couplings were made finite
in \cite{JPG-2020b}. The Abelian
contributions were found to be given by:
\begin{eqnarray} \label{Ifree}
{\cal L}_{free} =
- \frac{g_f^{(0)}}{4} 
\left( {\cal F}^{\mu\nu}_i {\cal F}_{\mu\nu}^i 
+ 
{\cal G}^{\mu\nu}_i {\cal G}_{\mu\nu}^i
+ 
{\cal F}^{\mu\nu} {\cal F}_{\mu\nu}
+
{\cal G}^{\mu\nu} {\cal G}_{\mu\nu}
\right),
\nonumber
\end{eqnarray}
where 
 the following effective parameter have been defined in the 
long wavelength and 
zero momentum limit considered before:
$ 
g_{f}^{(0)}   =
 d_1 4 N_c \; Tr' \; 
(( \tilde{S}_0^2 (k)))$.
This parameter can be set finite  as a renormalization condition, the one for
$Z_V$ and $Z_A$, and this
makes all the coupling constants and coupling functions in eq. (\ref{GVA}) finite.
The mass terms for the vector and axial mesons can also be found 
from this approach   with complementary contributions in similar 
developments \cite{curvature-mass-kovacs-etal,kim-lee-symmetric-vecmesmass}.
However the masses does not provide important information for 
the coupling constants and form factors addressed in the present work.
To obtain terms consistent with the massive Yang Mills approach, one must 
impose as renormalization point:
 $g_f^{(0)} = 1$.
 It follows that:
\begin{eqnarray}  \label{GVAPI-norm}
G_{v-a-\pi} (0,0) \sim  \frac{ g_{f}^{(0)} M^*}{2} 
+  \frac{1}{e}  g_{F\rho\omega} ,
\end{eqnarray}
where $g_{F\rho\omega}$ is the (finite) coupling constant of the  
of the neutral rho meson coupling to  a  photon and to the omega meson:
${\cal L}_{F\rho\omega} = - g_{F\rho\omega} 
F_{\mu\nu} {\cal F}^{\nu\rho}_3
{\cal F}_\rho^{ \mu}$ for $F_{\mu\nu}$ the background  photon 
stress tensor.

Finally,
there are other  three-leg couplings that can also be 
associated to mixing of $\rho$ and other vector and axial mesons, in particular 
the $\rho-\omega-A_1$ vertex 
\cite{sasaki-plb,kaiser-etal}. Also couplings that yield fusion of vector mesons into 
an axial mesons $g_{VVA}$.
It can be written as:
\begin{eqnarray}
{\cal L}_{3-v} &=& 
g_{\omega \rho A_1} (K,Q) 2 \epsilon^{\mu\nu\alpha\beta}
V_\mu (K+Q) \;  \left( {\cal F}_{\nu\alpha}^i (K)  A_\beta^i (Q)
+ {\cal G}_{\nu\alpha}^i (Q)  V_\beta^i (K) \right) + {\cal O}_{\cal F}
\nonumber
\\
&+& g_{\omega \omega f_1} (K,Q) 2 \epsilon^{\mu\nu\alpha\beta}
V_\mu (K+Q) \;  \left( {\cal F}_{\nu\alpha} (K)  A_\beta (Q)
+ {\cal G}_{\nu\alpha} (Q)  V_\beta (K) \right) + {\cal O}_{\cal F}
\nonumber
\\
&+& g_{\rho \rho A_1} (K,Q) 2 \epsilon^{\mu\nu\alpha\beta} i \epsilon_{ijk}
V_\mu^i (K+Q) \;  \left( {\cal F}_{\nu\alpha}^j (K)  A_\beta^k (Q)
+ {\cal G}_{\nu\alpha}^j (Q)  V_\beta^k (K) \right) + {\cal O}_{\cal F}
\end{eqnarray}
where ${\cal O}_{\cal F}$ contains a different momentum structures.
These other structures 
are obtained by exchanging the roles of the rho and the omega, or the $A_1$ meson,
i.e. ${\cal F} \leftrightarrow {\cal F}_i \leftrightarrow {\cal G}_i$
and correspondingly $V_\mu^i \leftrightarrow V_\mu \leftrightarrow A_{\mu}^i$.
The coupling function $g_{\omega\rho A_1}$, that is symmetrized due to the 
possible different order of
external lines, 
 was defined as:
\begin{eqnarray} \label{gora1-ren}
g_{\omega \rho A_1} (K,Q) &=&  C_3
\int_k  ((    S_0 (k) S_0 (k+K) \tilde{S}_0 (k+K+Q)
+   \tilde{S}_0 (k)  {S}_0 (k+K) {S}_0 (k+K+Q)  )),
\\ \label{gvvv}
g_{\omega \rho A_1} (K,Q) &=&   g_{\omega \omega f_1} (K,Q)
=  g_{\rho \rho A_1} (K,Q),
\end{eqnarray}
where $C_3=6 N_c d_2$, the 
momentum  integral has the same UV divergence as eq. (\ref{GVA-def}) and they can be
renormalized exactly in the same way as discussed above.
The right hand side ratio (\ref{gvvv}) is  also obtained from the naive quark model relation with 
a VMD hypothesis \cite{f1285prod}.
For constant quark effective mass, it can be written:
\begin{eqnarray} \label{gora1-vapi}
g_{\omega\rho A_1}  (Q,K) = \frac{1}{M^*} G_{v-a-\pi} (Q,K).
\end{eqnarray}
These coupling constants  are also directly proportional to 
coupling constants found in the framework of the Skyrme model 
or massive Yang Mills model
 \cite{skyrme,vecmes,hohler-rapp}.

\subsection{ Possible quantization of meson couplings to axial and vector   couplings
}
\label{sec:quantization}

The coupling  $F_{vja} (K,Q)$ in the action  can be written in the 
coordinate space and,
with 
 the same construction of Witten \cite{witten-1983,weinberg-book},
it can be expressed as
 a five dimensional closed  surface of a total divergence
by means of the Stoke's theorem.
In this case
a  quantization condition emerges.
By writing it back 
in the momentum space,  eq. (\ref{L-vja})
can be written as:
\begin{eqnarray} \label{quant}
n \Gamma 
= -  \frac{i}{240 \pi^2 } 
\int d^4 K \;  d^4 Q \;
\epsilon_{\sigma\rho\mu\nu } 
 F^{vja} (K,Q)  
K_\sigma {\cal F}^i_{\rho \mu} (Q) 
j^{A, i}_\nu  (K,K+Q) ,
\end{eqnarray}
where $n$ is an integer.
This integral, as a term in the action, 
 corresponds therefore to a topologically conserved quantity
assuming integer multiple values  of $\Gamma$.
The axial current however  has specific properties.
For instance,  the
axial charge  is only partially  conserved due to both 
quark masses and the ABJ anomaly.
So the following question arise: 
how can a topologically conserved coupling involving the axial current
be related to the partially conservation  of the  axial current?
Seemingly the vector meson anomalous coupling selects a component of $j_\nu^{A,i}$ that 
would be quantized and conserved.
By decomposing the above integral into the different Lorentz
components, i.e. $n \Gamma  = \epsilon^{\sigma\rho\mu\nu} \Gamma_{(\sigma\rho\mu\nu)}$,
one can write explicitly one of these terms as:
\begin{eqnarray} \label{quant-KQ}
 \Gamma_{(xyz0)} 
&=& 
 -  \frac{i}{240 \pi^2 } 
\int d^4 K \; d^4 Q \;
F^{vja} (K,Q) K_x Q_y 
\nonumber
\\
\times
&&
\left[ \rho^{-}_z (Q)  \bar{u} (K+Q) \gamma_0\gamma_5 d (K)
+  \rho^{+}_z (Q)  \bar{d} (K+Q)  \gamma_0\gamma_5 u (K) \right],
\end{eqnarray}
where $\rho^\pm_z (Q)$ are the z-polarization component profile 
of the 
charged rho field.
The quantization condition however involves the sum of 
all the couplings of all the components
$\Gamma_{(xyz0)}$.

Besides that, for constant effective mass,
 these equations can be rewritten in terms of the 
pion axial coupling to constituent quarks, eq.  (\ref{GA-U}).
From  eq. (\ref{ratio-GA-FVJA}) it can be written that:
\begin{eqnarray} \label{gamma-gA}
n \Gamma =
-  \frac{i}{ 240 \pi^2 \times 4 M^* F}
\int d^4 K \; d^4 Q \; 
\epsilon^{\sigma \rho \mu \nu}
G_A (K,Q) 
K_\sigma {\cal F}^i_{\rho \mu} (Q) 
j^{A, j}_\nu  (K,K+Q) .
\end{eqnarray}
Note however, that the momenta $K$ and $Q$  are orthogonal to each other,
differently from the pion axial coupling to the constituent quark current.
This eq. (\ref{gamma-gA}) has a higher order dependence on 
the constituent quark momentum $K_\sigma$
but it is of the same order of the meson  momentum (pion or vector meson) $Q_\rho$.

By applying the same reasoning to the
axial meson coupling to the constituent quark vector current (\ref{L-vja-A})
the following quantization conditions are obtained:
\begin{eqnarray}
\label{quantiz-vja-V} 
m_t \bar\Gamma
&=&
\int d^4 x_1 d^4 x_2 \;
 i  \epsilon^{\sigma \rho \mu  \nu}
  F^{vja} (x_1,x_2)  
 {\cal G}^i_{\rho \mu} (x_1) 
\partial_\sigma j^{V, i}_\nu  (x_2,x_1),
\\  \label{quantiz-vja-V-s} 
m_s \tilde\Gamma
&=&
\int d^4 x_1 d^4 x_2 \;
 i  \epsilon^{\sigma \rho \mu  \nu}
  F^{vja} (x_1,x_2)  
 {\cal G}_{\rho \mu} (x_1) 
\partial_\sigma j^{V}_\nu  (x_2,x_1) 
,
\end{eqnarray}
where $m_s, m_t$ are integers for the isosinglet and isotriplet 
axial mesons interactions.
Notwithstanding the vector current  is conserved only  for  
degenerate quark masses
these conditions
should not disappear in the case of non
degeneracy of quark masses.
A similar question to the one raised above arises: 
how can a topologically conserved quantity, involving the
constituent quark vector current,  be related to 
the non conservation of the (global)  vector current that is due to 
the non degenerate quark masses?
Similarly to the axial current coupling to the vector meson,
the axial meson profile and coupling might 
"select" a topologically preserved component of the vector current.
The effect of quark mass non-degeneracy will be investigated in another
work.

\section{ Anomalous vector mesons  couplings 
to the quark axial current under weak magnetic field}
\label{sec:Bfield}

The effect of a  magnetic  field,
weak with respect to the constituent quark mass $M^*$, will be
presented due to two different mechanisms
described in detail  in Ref. \cite{JPG-2020a}.
 The validity of the (semi)classical approximation for the  magnetic field to describe 
observables in heavy ion collisions has been object of attention in the last years 
\cite{navarra+est,navarra+est2}.
The  dependence of the  quark propagator 
 on the weak magnetic field 
is considered and, secondly, the overall photon couplings to the legs of the 
meson-current coupling.
For the degenerate quark effective mass $M^*$, 
the quark propagator,  with the leading contribution 
from the weak magnetic field  aligned in the $z$-direction,
  can be written for degenerate quark masses 
as  \cite{weak-B,igor1}:
\begin{eqnarray}  \label{quark-prop}
G (k) &=&  S_0 (k) + S_1 (k) (e B_0) = 
\frac{ \slashed{k}+  {M^*} }{ k^2 - {M^*}^2 + i\epsilon } 
+ i \gamma_1 \gamma_2
 \frac{
 (  \gamma_0 k^0 - \gamma_3 k^3  +  {M^*}  )
 }{
(k^2 - {M^*}^2 + i \epsilon )^2 }  \hat{Q} (e B_0) .
\end{eqnarray}

The   anomalous light vector mesons couplings 
to the axial current  
in the presence of   this  constant weak
magnetic field 
provide similar expressions 
in both mechanisms mentioned above 
that 
can be added.
Below some of  resulting corrections for the couplings, with the form factors, 
of the previous section are shown
 for the isotriplet and isosinglet vector and axial mesons.
By writing them with a  constant small 
multiplicative factor $(eB_0)/{M^*}^2$,
they can be  written as: 
\begin{eqnarray} \label{L-VJA-B}
{\cal L}_{vjaB}  &=& 
\frac{(eB_0)}{{M^*}^2} 
\epsilon_{ij3}
 \frac{F_{vja}^{B} (K,Q) }{{M^*}^2}   \left[
 \epsilon^{12\rho\mu}   K_\rho  Q_\nu \cdot V_\nu^i (Q)  
+  
2  \epsilon_{12\rho\nu}  K^\rho {\cal F}^{\mu\nu}_i  (Q)
 \right] 
\; j^{A, j}_\mu  (K,K+Q) 
\nonumber
\\
&+& 
\frac{(eB_0)}{{M^*}^2} 
 \frac{F_{vja}^{B} (K,Q) }{3 {M^*}^2 } \left[
 \epsilon^{12\rho\mu}   K_\rho  Q_\nu \cdot V_\nu (Q)  
+  
2  \epsilon_{12\rho\nu}  K^\rho {\cal F}^{\mu\nu}  (Q)
 \right] 
\; j^{A, 3}_\mu  (K,K+Q) ,
\\ \label{mix-B}
{\cal L}_{mix,B}
&=&
\frac{(eB_0)}{{M^*}^2} 
G_{v-a}^{B,1} (K,Q)  i \epsilon_{12\mu\nu} 
{M^*}^2 
i \epsilon_{ij3}  V^\mu_i (Q)  \bar{A}^\nu_j (K) \delta (Q+K)
\\
&+&
\frac{(eB_0)}{{M^*}^2} 
G_{v-a-\pi}^{B,1} (Q,K)  i \epsilon_{12\mu\nu} 
{M^*} \left(
\pi_3 (Q) V^\mu (K) \bar{A}^\nu (Q+K)
+ T_{ijk} \pi_i (Q) V^\mu_j (K) \bar{A}^\nu_k (K+Q)
\right)
,
\nonumber
\end{eqnarray}
where 
 $T_{ijk} = 
tr_F  (Q \sigma_i \sigma_j \sigma_k)$.
The coupling functions
can be written in the Euclidean momentum space  as:
\begin{eqnarray}
 F_{vja}^{B}   (K,Q)  
 &=& 
 4 d_2 N_c   (\alpha g^2)  {M^*}^4  \; \int_k  \;     
((   \tilde{S}_0 (k+K) 
 \tilde{S}_0 (k+K) \tilde{S}_0 (k+K+Q) \bar{R} (-k)  )) ,
\\
G_{v-a}^{B,1} (K,Q) \delta(K+Q)   &=&  2  d_2 N_c  
 \; \int_k  ((  {M^*}^2 \tilde{S}_0  (k+Q)  \tilde{S}_0  (k+Q) \tilde{S}_0 (k) )),
\\
G_{v-a-\pi}^{B,1} (Q,K) &=&
3  d_2 N_c   {M^*}^2
\\
&& \times 
 \; \int_k  ((   ( k\cdot (k+Q) - {M^*}^2) 
 \tilde{S}_0  (k+Q)  \tilde{S}_0  (k+Q) \tilde{S}_0 (k+K+Q) 
\tilde{S}_0(k) )),
 \nonumber
\end{eqnarray}
where    the vector mesons 
 are the canonically normalized ones.
These coupling functions (coupling constants) are all finite
and they are dimensionless.
The couplings $F_{vja}^B(K,Q)$ might be seen as 
magnetic field corrections to the anomalous vector meson form factor.

 The
mixing $G_{v-a}^{B,1} (K,Q)$
 disappears in the absence of magnetic field.
Note that the similar term for isosinglet vector and axial fields 
does not show up.
This term, however,  is trivially  zero in the absence of other particles
due to conservation of momentum, and this 
is made  explicit in the delta function $\delta (K+Q)$.
In the presence of another particle, or in a finite density medium, 
the conservation of  momentum can
be satisfied and  this anomalous mixing can contribute 
because
  vector and axial mesons propagate in orthogonal directions.
This coupling should  contribute for in medium vector-axial
mixing besides the  mixing induced by the pion of eq. (\ref{GVA})
\cite{rho-a1-1,harada-etal}.
All these interaction terms
  emerge due to the anomalous trace
of five Dirac matrices, eq. (\ref{trace=5Dirac}).
There is  a similar  momentum anisotropy to 
 the case of zero magnetic field, being also dependent  on the vector
or axial  mesons polarization.
 The magnetic field contributions are suppressed by factors $1/{M^*}$ or $1/{M^*}^2$
with respect to the zero magnetic field couplings,
besides the  factors $(eB_0)/{M^*}^2$ were included to.
Although the explicit magnetic field contribution is factorized and suppressed
by a factor $(eB_0)/{M^*}^2$,
the effective quark masses also depend on the magnetic field 
in  the gap equations.
These anomalous form factors  
can be simply added to
the anomalous form factors (\ref{L-vja},\ref{L-vja-A}) and (\ref{GVA})
 in the vacuum as it follows:
\begin{eqnarray} \label{F+FB}
F (K,Q ; B_0) = 
F (K,Q) + f \frac{(eB_0)}{{M^*}^2} F^B (K,Q),
\end{eqnarray}
where $f$ is a factor that depends on the momentum components
in eq. (\ref{L-VJA-B}) or (\ref{mix-B}).
 $F_{vja}(K,Q)$ and $F_{vja}^B (K,Q)$ 
also can
receive magnetic field corrections from the gluon propagator and running 
coupling constants dependencies on $B_0$ being however 
 neglected  in the 
present work. 
Axial vector mesons couplings to the quark vector currents,
eq. (\ref{L-vja-A}),
receive analogous corrections 
 due to 
the weak magnetic field to those shown above.

\section{ Numerical results}
\label{sec:numerical}

For the numerical estimations
two different effective gluon propagators will be considered.
{
The first  one will be a
transversal
 obtained from Schwinger-Dyson equations
(SDE) hat   reproduce  many hadron observables 
\cite{maris-craig-tandy,SD-rainbow}.
It can be written as:
\begin{eqnarray} \label{gluon-prop-sde}
 D_{I} (k)  = h_{I} g^2 R_T (k)  &=&  
\frac{8  \pi^2}{\omega^4} De^{-k^2/\omega^2}
+ \frac{8 \pi^2 \gamma_m E(k^2)}{ \ln
 \left[ \tau + ( 1 + k^2/\Lambda^2_{QCD} )^2 
\right]}
,
\end{eqnarray}
where {$h_I$ is the factor that normalizes the coupling constant of the 
vector meson 
to the vector current,
$\gamma_m=12/(33-2N_f)$, $N_f=4$, $\Lambda_{QCD}=0.234$GeV,
$\tau=e^2-1$, $E(k^2)=[ 1- exp(-k^2/[4m_t^2])/k^2$, $m_t=0.5 GeV$,
$D= 0.55^3/\omega$ (GeV$^2$) and 
$\omega = 0.5$GeV.

The second type  
 is based in a longitudinal  effective confining parameterization
 by Cornwall \cite{cornwall}
that can be  written  as:
\begin{eqnarray} \label{cornwall}
R_L (k) = D_{II,z} (k) &=& h_{II,z}
\frac{ K_F }{(k^2+ M_z^2)^2} ,
\end{eqnarray} 
 where 
$K_F = (2 \pi M_z/3 k_e)^2$ was  considered in previous works
\cite{PRD-2018a,PRD-2019}.
In this eq., $k_e\simeq 0.15$, 
the  normalization factor $h_{II,k}$  is
 considered to reproduce a reasonable value for 
the vector meson (rho or omega) coupling constant to the nucleon
(or constituent quarks). 
This type of gluon effective propagator exhibits features 
of
 lattice QCD calculations \cite{ooliveira-etal}.
Two  choices were made for the   effective gluon mass $M_z^2$: 
as a function of 
momentum  ($z=6$ and $z=7$) for 
$M_6 = 0.5/(1+ k^2)$ GeV
and $M_7 = 0.5/(1+ k^2/5)$ GeV.

In the figures below the following 
 normalized space-like 
form factors with respect to the usual vector meson coupling to the vector current
 will be drawn:}
\begin{eqnarray}   \label{F-G-vja} 
G_{vja} (K,Q) = \frac{  F_{vja} (K,Q)}{  G_{V} (0,0) } ,
\;\;\;\;\;\;\;
G_{vja}^B (K,Q) = \frac{  F_{vja}^B (K,Q)}{  G_{V} (0,0) } ,
\end{eqnarray} 
 where the following normalizing parameters where chosen:
$h_I = 1.4$ ($M^*=0.33$GeV),
$h_I = 1.6$ ($M^*=0.45$GeV),
 and $h_{II,6} = h_{II,7} = 16$ ($M^*=0.33$GeV).
These values lead to $G_V(0,0) = 12$ \cite{PRD-2018a}
This way we can extract  relative strengths of the couplings
 in an uniform  way to  make comparisons.
Besides that, only one component of $K$ and $Q$  contribute, and, due to this,
we considered (in a spacelike spherical coordinate basis)
 $Q = |Q|/4$ and $K  \simeq |K|/4$.

In Figure (\ref{fig:Gvja-TM})
the form factor 
$G_{vja} (K,Q)$,
calculated with the effective gluon propagator
$D_I(k)$, is presented as a function  
of the  vector meson momentum $Q^2$
 for 
different values of the 
modulus of the constituent quark momentum $K=  |K|/4$.
Two different values of the quark 
effective mass, typical from the NJL-model,
 are considered  $M^*=0.33$GeV and $M^*=0.45$ GeV.
The momentum dependencies in $K$ and $Q$ are not very different, 
as it can be seen in 
eq. (\ref{Gvja}).
There is a slight increase  up to   $Q^2 \sim (0.2GeV)^2$ or $K^2\sim (0.2GeV)^2$
and  then a decrease with both $Q^2$ and $K^2$.
For the zero or very low quark  momenta $K\sim0$ limit the behavior with $Q^2$ is 
almost monotonically decreasing.
A larger quark effective mass leads to a suppression of the interaction
and to a less strong dependence on momenta.
This  figure, and the following ones,  
show that there might have a weak
 coupling of the light vector mesons, $\rho$ and/or $\omega$,
 with the axial constituent quark current.
The relative strength of the coupling $F_{vja}(K,Q)$ and the usual 
vector coupling   close to $Q \sim K \sim 200-500$MeV
 can be estimated to be:
\begin{eqnarray}
\left. \frac{F_{vja}(K,Q) \times |K| |Q|}{ G_V (K,Q) } \right|_{Q \sim K \sim 200 -500 MeV}
  \sim  0.1 .
\end{eqnarray}
This coupling function also provides an 
(anomalous) axial contribution for the vector meson form factor.
Therefore only precision measurements of the vector meson interactions and form factors 
would identify such contributions.
It is important to emphasize that $F_{vja}(K,Q)$  is 
 strongly anisotropic, being that
it selects a particular meson polarization.
Furthermore, $F_{vja}(K,Q)$  for larger momenta is suppressed stronger 
than the vector coupling $G_V(K,Q)$.
Its (anisotropic) effects could also be searched in the
strength of the vector meson dominance (VMD)
\cite{nucleon-eff}.
Besides that, these couplings correspond to anomalous axial corrections
to the rho or omega ($A_1$ or $f_1$) form factors
for which there are some estimations \cite{ballon-etal,krutov-etal,PRD-2018a}.

\begin{figure}[ht!]
\centering
\includegraphics[width=120mm]{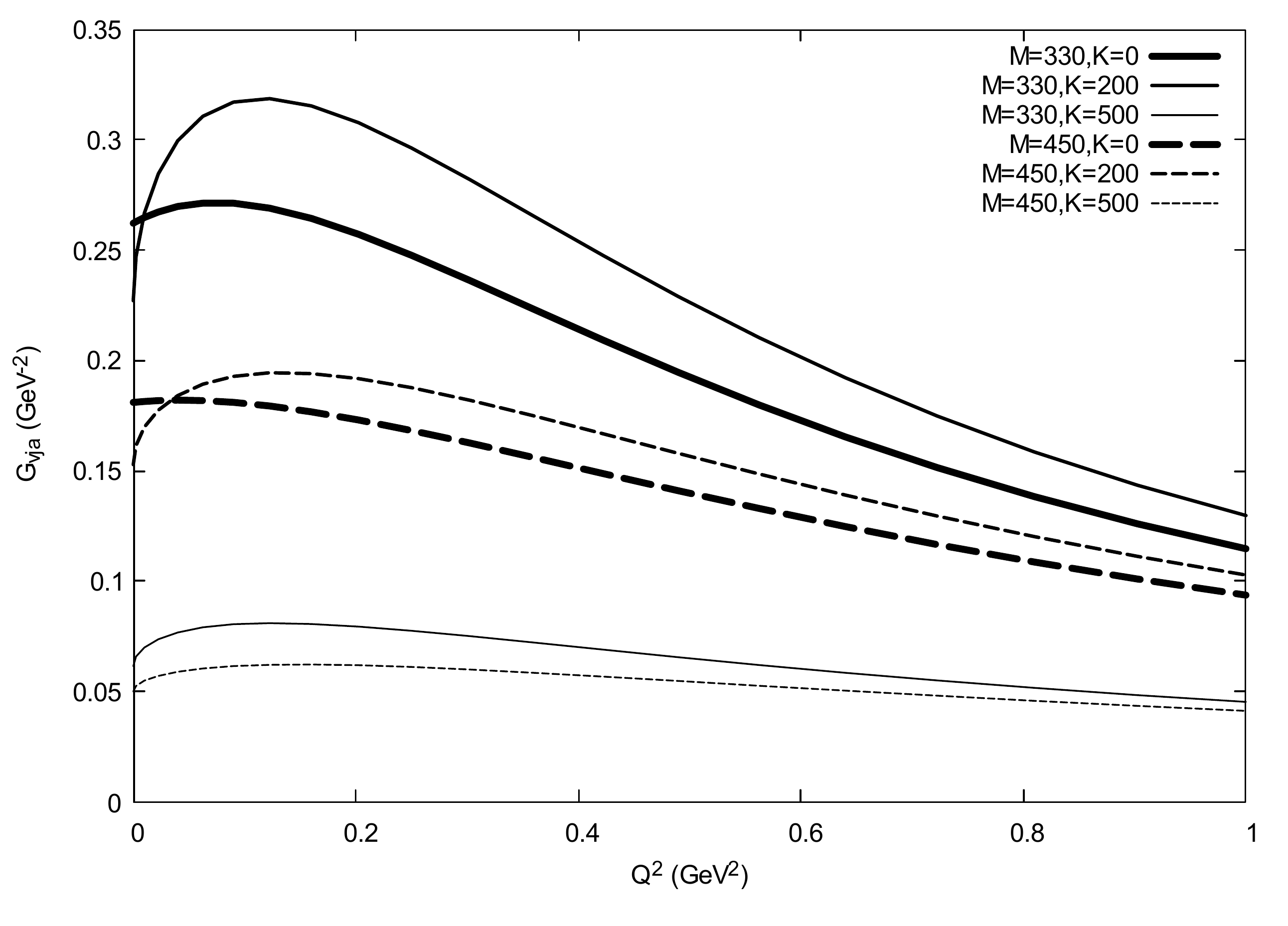}
 \caption{
\small
Anomalous   form factor  $G_{vja} (K,Q)$
for  the  effective
  gluon propagator $D_I(k)$ as a function 
 $Q^2$ for 
different values  of  $K$.
Two different quark effective masses are used
$M^*=0.33$GeV and $M^*=0.45$ GeV.
}
 \label{fig:Gvja-TM}
\end{figure}
\FloatBarrier

The same  form factor $G_{vja}(K,Q)$ as a function of $Q^2$ 
is exhibited, for different values of $K=0, 200, 500$MeV,
for the effective gluon propagators $D_{II,6}(k)$ and $D_{II,7}(k)$
in figure (\ref{fig:Gvja-CO67}) for the quark effective mass $M^*=0.33$GeV.
The behavior is very similar to the one found in figure (\ref{fig:Gvja-TM}) for 
the gluon propagator $D_{I}(k)$.
The very small difference between the results from the  gluon propagators only shows up 
for low-intermediary momenta, and it tends to zero  for higher momenta
 $K > 400$MeV and $Q^2 > 1$GeV$^{2}$.

\begin{figure}[ht!]
\centering
\includegraphics[width=120mm]{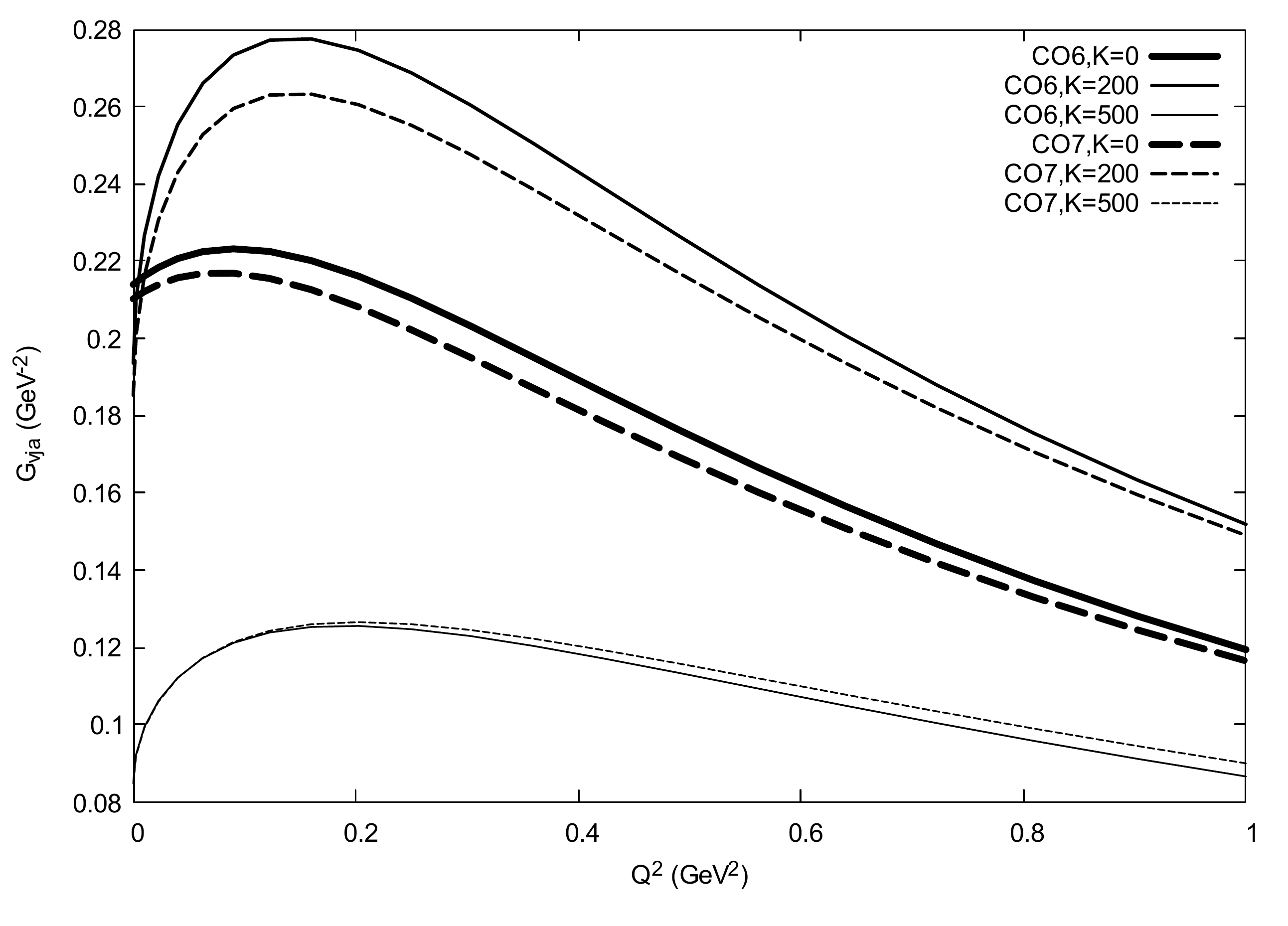}
 \caption{ 
\small
Anomalous axial  form factors  $G_{vja} (K,Q)$
for  the two effective
  gluon propagators $D_6(k)$ 
and $D_{7}(k)$,  and
with 
$M^*=0.33$ GeV.
}
\label{fig:Gvja-CO67}
\end{figure}
\FloatBarrier

\subsection{ Magnetic field induced correction}

In figure (\ref{fig:GvjaB-TM}) 
the  form factor $G_{vja}^B (K,Q)$ 
 is drawn for the gluon propagator $D_I(k)$ 
and two different quark effective masses, $M^*=0.33$GeV and $0.45$GeV,
 as a function
of the vector meson momenta $Q^2$ for the same 
quark momentum as the previous figures.
The effect of modifying the quark effective mass is much larger 
 at low momenta $Q$ and $K$. For higher momenta, 
 the difference 
between the form factors with the different effective masses 
tends to become
 smaller with $Q^2$ and  with $K^2$.
Also, analogously to the zero magnetic field case exhibited above, 
 there is a small increase with low momenta up to 
$K \sim 200$MeV or $Q\sim 200$MeV.
In figure (\ref{fig:GvjaB-CO67})
the same form factor  $G_{vja}^B  (K,Q)$ 
is presented 
for the same  values of $K=0, 200, 500$MeV, with $M^*=0.33$ GeV and 
for the effective gluon propagators $D_{II,6}(k)$ and $D_{II,7}(k)$.
The behavior with external momenta is basically the same as the form factor $G_{vja}$
and the difference between results of the two different gluon propagators
shows up in intermediary momenta.
This  anomalous  constituent quark and vector meson   coupling function
could manifest in the low/intermediary vector meson energy regime
in ultra-peripheral collisions.

\begin{figure}[ht!]
\centering
\includegraphics[width=120mm]{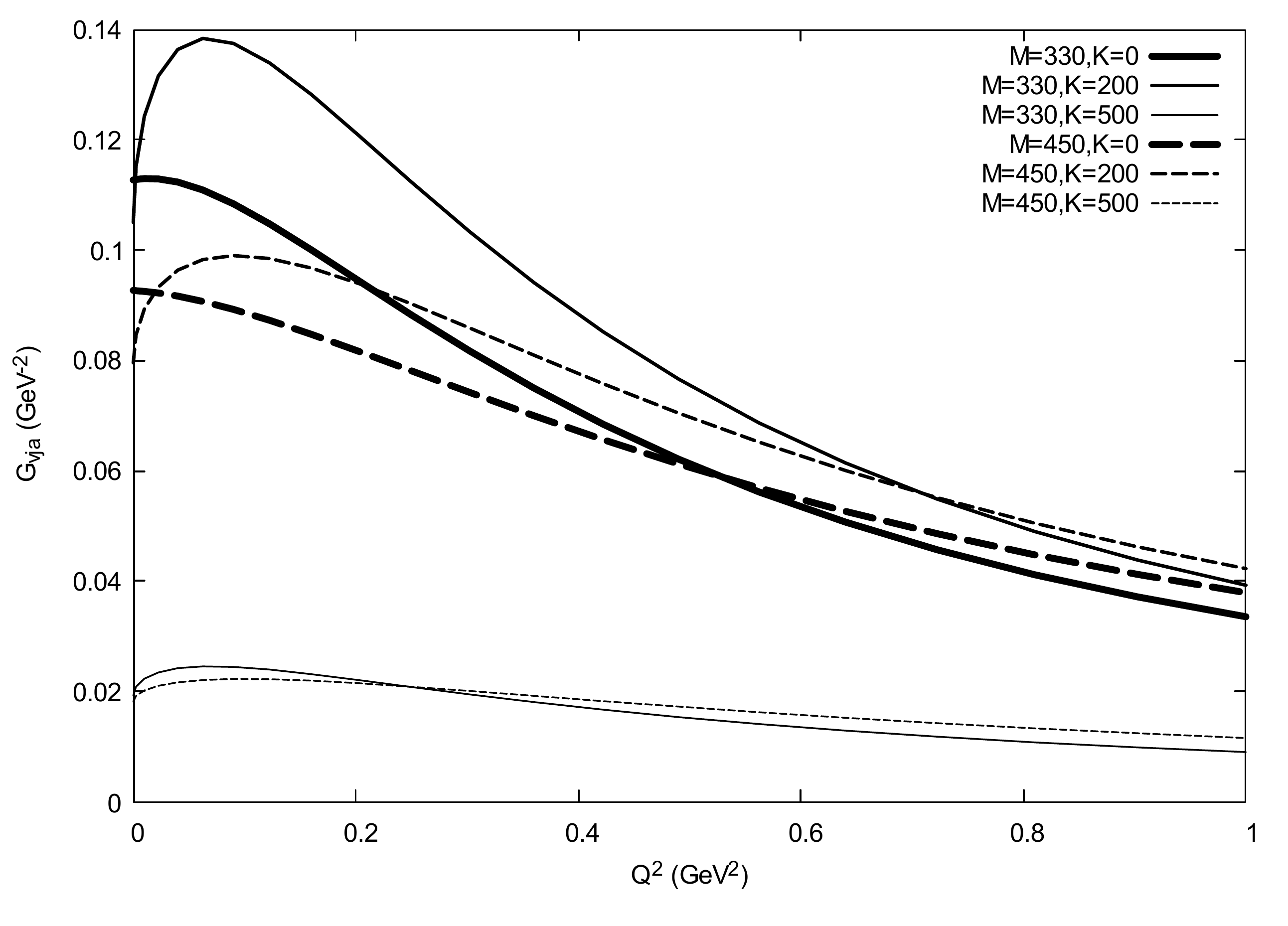}
 \caption{ \label{fig:GvjaB-TM}
\small
 Magnetic field induced  correction to the  anomalous axial  form factors 
probed by the rho vector meson
for  the effective
  gluon propagator $D_I(k)$ 
and for two different quark effective masses
$M^*=0.33$GeV and $M^*=0.450$ GeV.
Modulus of quark momentum was chosen to be the same of the previous figures:
$K = 0, 200, 500$MeV.
}
\end{figure}
\FloatBarrier

\begin{figure}[ht!]
\centering
\includegraphics[width=120mm]{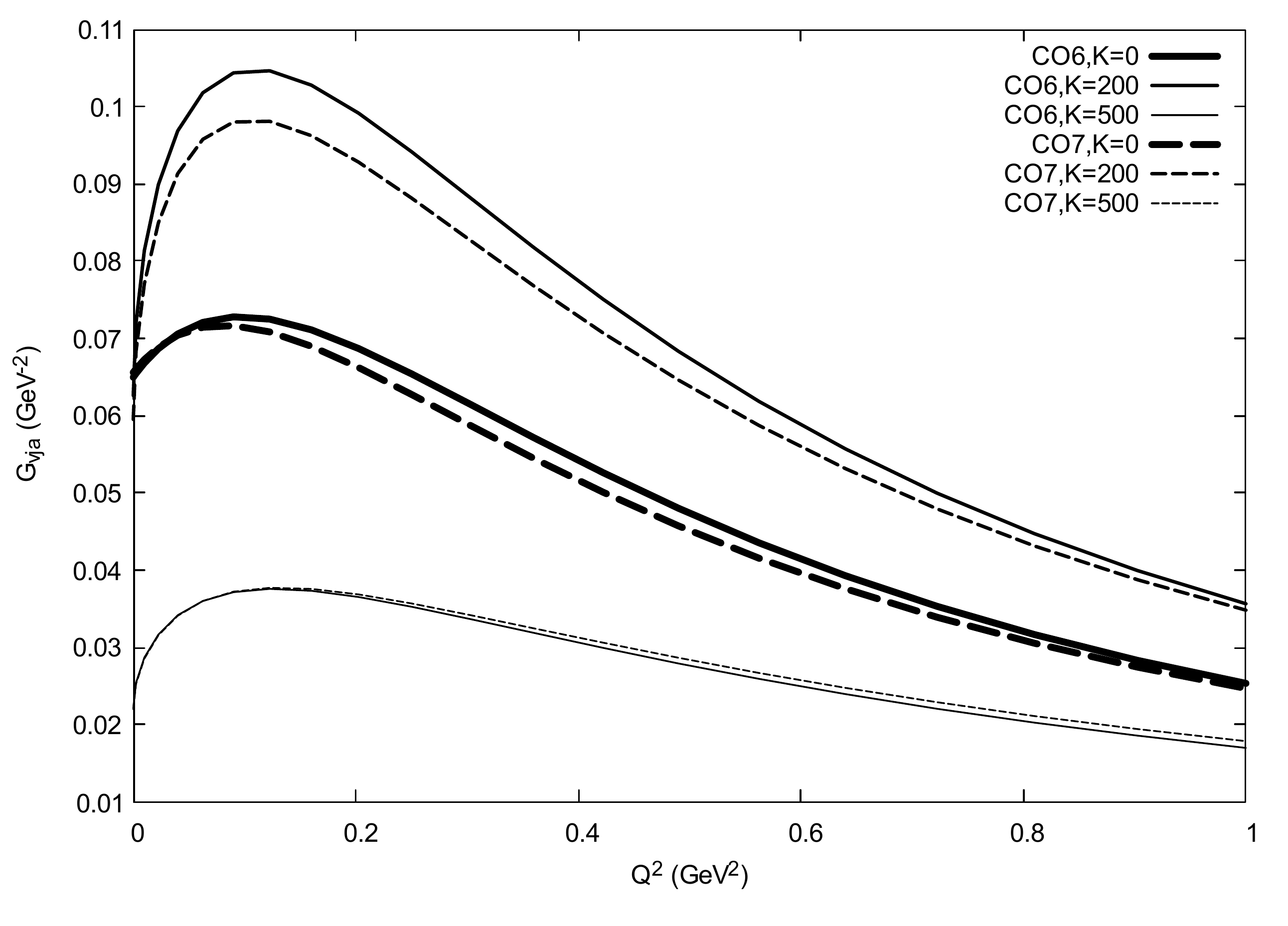}
 \caption{ \label{fig:GvjaB-CO67}
\small
The same as figure (\ref{fig:GvjaB-TM})
for  the two effective
  gluon propagators $D_6(k)$ 
and $D_{7}(k)$, 
with  
$M^*=0.33$ GeV 
and $K = 0, 200, 500$ MeV.
}
\end{figure}
\FloatBarrier

\subsection{
 Averaged quadratic radius
and  on-shell estimates for coupling constants}

Vector (and axial) mesons   couple on shell although
 constituent quarks cannot  be really said to do so.
Below some estimates are provided for  coupling constants
with  on shell vector mesons.
Because of the involved 
analytic structure,
 only gluon effective propagator (\ref{cornwall})
will be considered
 for which we adopted the  value $M_6=500$ MeV.
The momentum integral of the coupling constant $g_{\omega\rho A_1}$ is ultraviolet divergent.
Because of that, a regularized truncated version of the 
quark propagator was considered  \cite{PRD-2018a,JPG-2020a}:
$S_0(k) \to  M^*/(k^2 - {M^*}^2)$.
It may yield an effective momentum dependence close to the
ones obtained from SDE.
With that, the integral becomes finite
and we denote the result by
$G_{\omega\rho A_1}^{tr} (K,Q)$.
All the singularities were taken into account
by considering the average of the integration over the up and down complex semi-planes.
The following ranges of values were obtained for a range of quark effective masses
$M^*=0.33-0.35$  GeV:
\begin{eqnarray}
  G_{vja} (K_0=M^*, Q_0 =M_\rho) &=& 0.6 - 0.4 \; GeV^{-2},
\\
 G_{\omega\rho A_1}^{tr} (K_0=M_\rho, Q_0=M_\omega) &=& 0.2-0.3 , \;\;\;\;
\\
G_{kv-a-\pi} (K_0=M_\rho, Q_0=M_{A_1}) &=&    0.9 - 1.2 \; GeV^{-1}, \;\;\;\;
\\
 G_{v-a}^{B_1} (K_0=M_\rho)  &=& 0.1-0.3,
\end{eqnarray}
Note that: $G_{v-a-\pi}^{tr}  = {M^*}^2 G_{kv-a-\pi}$.
Also, from eq. (\ref{gvvv}), one has for the vector mesons fusion channel:
$g_{\omega \rho A_1} (K_0,Q_0) \simeq  g_{\omega \omega f_1} (K_0,Q_0)
\simeq  g_{\rho \rho A_1} (K_0,Q_0)$, since the 
mass difference $M_\rho - M_\omega \sim 12$MeV is very small

For the sake of comparison,  few values  for the coupling $G_{\omega\rho A_1}$ 
or $G_{VA\pi}$ considered in the literature
are quoted next.
The coupling constant $G_{\omega\rho A_1}$ has been considered 
 for the investigations 
(of vector/axial  mesons mixing) at finite density  
\cite{sasaki-plb,rho-a1-1}. 
The following values were used:
$g_{\omega \rho A_1} <\omega_0>= C
 \simeq 0.1 \to 0.3$  GeV, at the saturation density $\rho_0$
where
the quantity  $<\omega_0>$  is the omega mean field in the medium.
From Ref. \cite{du-zhao} the following values were considered for the dimensionless 
three-leg coupling given in (\ref{gora1-vapi}):
 $ G_{kv-a-\pi} \sim M^* G_{AVP} \sim 2 M^* \sim 0.7$.

A simply estimation for the contribution of the form factor $F_{vja}(K,Q)$ to the 
rho meson averaged quadratic radius can be also provided
It corresponds to a small axial component.
For that, we can define a normalized dimensionless coupling function (form factor) as:
$\bar{G}_{vja} =   \bar{K} \bar{Q} G_{vja} (K,Q)$ where
$\bar{K}, \bar{Q} \sim 200$ MeV are averaged constituent quark  and meson momenta.
The usual definition of averaged quadratic radius (a.q.r.) will be adopted:
\begin{eqnarray} \label{rqm}
\Delta_A < r^2_\rho > = - 6 \; \left. \frac{ d \bar{G}_{vja} }{d Q^2} \right|_{Q=0}.
\end{eqnarray}
Results are shown in fig. (\ref{fig:fig5}) for two gluon propagators
$D_I$ and $D_{II,6}$ with $M_6=500$ MeV.
These values  can be compared to estimations of the 
rho a.q.r.:  $< r^2_\rho > \simeq 0.28-0.56$fm$^{2}$
\cite{PRD-2018a,ballon-etal,krutov-etal,bhagwat-maris,hlroberts-etal}.
It very small, i.e. $\sqrt{\Delta_A <r^2_\rho>}$ can be  one order of magnitude smaller
than the rho vector meson radius.
The contribution of the weak magnetic field is very small, it is 
one order of magnitude smaller than 
the above contribution, i.e.
$\Delta_A^B < r_\rho^2> \sim \frac{\Delta_A < r_\rho^2>}{10}$.

\begin{figure}[ht!]
\centering
\includegraphics[width=120mm]{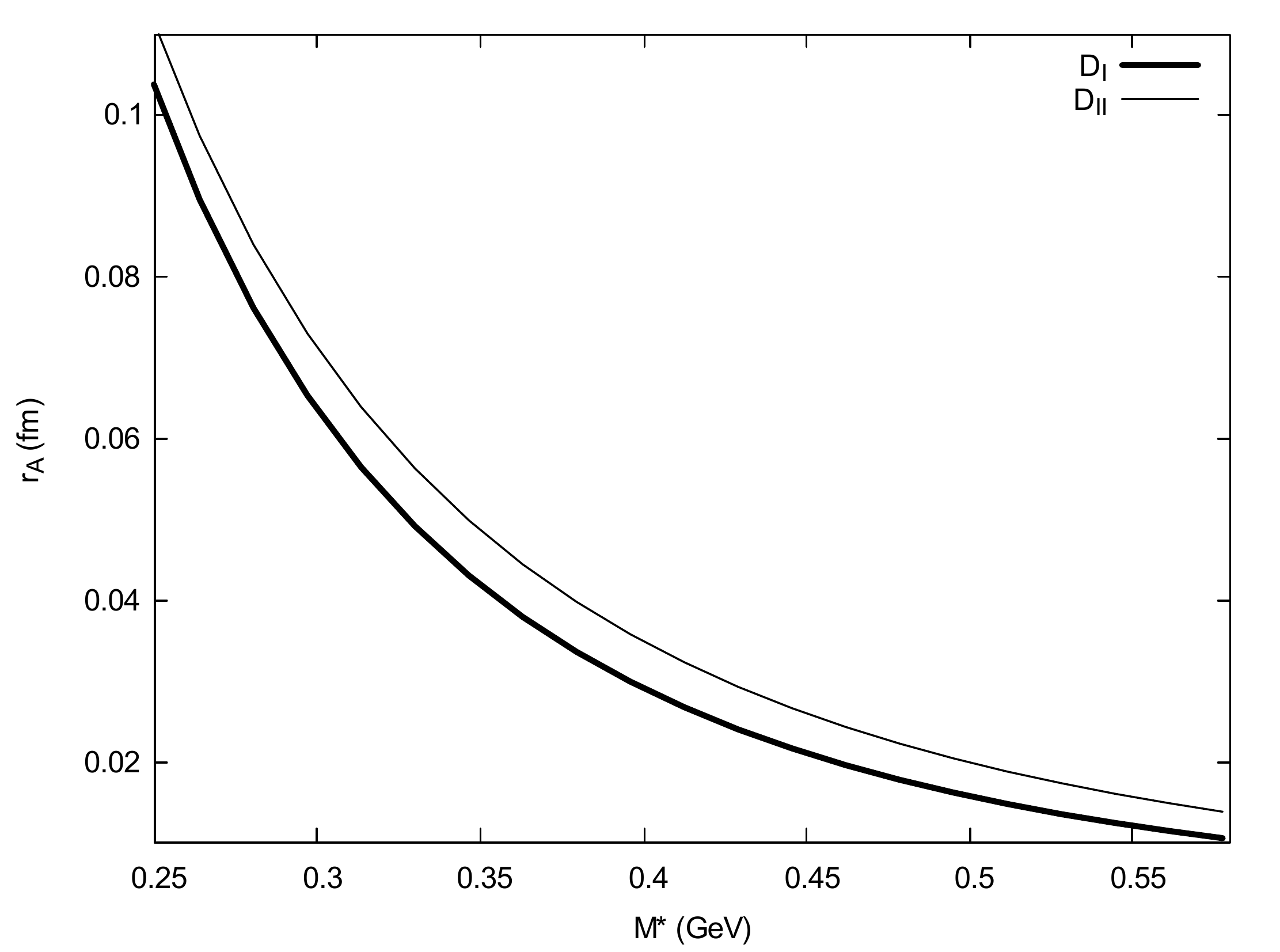}
 \caption{ \label{fig:fig5}
\small
The axial a.q.r., $r_A \equiv \sqrt{\Delta_A <r^2_\rho>}$,  obtained 
with $D_I$ and $D_{II,6}$ (with $M_6=500$MeV)
as a function of the quark effective mass.
}
\end{figure}
\FloatBarrier

\section{ Final remarks}

Unusual   couplings
  of    light vector mesons, rho and omega, to the 
constituent quark axial current  were derived in this work.
Their momentum dependencies were presented for different 
effective gluon propagators  in 
 the low momentum regime.
 The resulting coupling constants and form factors 
are  suppressed  at low momenta by 
 $(K_\mu Q_\nu)/{M^*}^2$
 in comparison with 
the minimal   vector mesons couplings to the quark vector current
 \cite{PRD-2018a}.
$K_\mu$ and  $Q_\nu$ are respectively the quark and vector meson momenta,
being  that the coupling functions are  strongly anisotropic.
The momentum dependencies of $F_{vja}$ on $Q$ and $K$
are basically the same.
The couplings are also   dependent on the  polarization of the vector meson.
Although they go to zero for very large, infinite, meson momenta, the resulting equations 
are suppressed fast with respect to the leading form factors at  higher energies.
Besides that,  at higher energies, quark effective masses should decrease and 
this might, to some extent,  invalidate the large quark mass expansion.
The aim of the present work is to provide a dynamical derivation for the low
energies mesons couplings.
As such, the resulting equations can be written in terms of structureless (local) meson fields.
When analyzing  topological-based equations from section (\ref{sec:quantization})
there is a need to integrated over all the meson field profile, or 
their momentum dependencies.
However, since  the structure of $F_{vja}(K,Q)$
decreases quite strongly with increasing momenta, 
the overall high energy contributions from the mesons profiles  should be suppressed.
This was not investigated further in the present work.

Numerically we have found $G_{vja}=F_{vja}/G_V \sim 0.1 - 0.3$ GeV$^{-2}$, i.e.
this coupling might produce small (anisotropic) effects in observables
associated to the vector meson coupling to constituent quarks or to  the nucleon.
The corresponding light axial mesons couplings to the constituent quark vector
current were also derived and they have the same strength
and overall momentum dependence.
When comparing the behavior of the form factors  $F_{vja}(K,Q)$, 
calculated with different 
effective gluon propagators, 
very small (negligible)  differences  were found mostly for momenta
0.15 GeV$^2 \leq Q^2 \leq$ 0.7 GeV$^{2}$.
Modifications in the value of the quark effective masses
lead to shifts in the low momenta region of $F_{vja}(K,Q)$ and 
to its weak magnetic field induced correction
$F_{vja}^B(K,Q)$.
These form factors correspond to axial (vector) components of 
the vector (axial) mesons.
Correspondingly a simple estimation
of the  axial  averaged quadratic radius for the rho meson was provided.
It is, at most,  one order of magnitude smaller than the estimations for  the 
rho a.q.r.
There might appear other consequences.
 For example,
a small (anisotropic) contribution for the
 so-called vector meson dominance
(VMD) and related form factors
\cite{nucleon-eff}.

Other   three-leg meson vertices often considered,
such as   $\rho-A_1-\pi$, $\rho\rho-A_1$ and $\omega-\rho-A_1$,
 were
 also derived  being the estimated values in quite good agreement with other investigations.
These three-meson coupling constants, or coupling functions,
 might be finite or
logarithmic  UV divergent. 
In this last case, they are renormalized  by the same renormalization constants as the 
free vector mesons kinetic  terms \cite{JPG-2020b}.
Finally, mixings between 
light vector and  axial mesons
were also found.
However the corresponding anomalous  mixing only  can  provide dynamical contributions
in the presence of a third particle or in a finite density medium.
 This is due to  the conservation of linear momentum. 
Incoming (or outgoing) vector meson and outgoing (or incoming) axial meson
must  propagate in orthogonal directions.
These couplings and  mixings are    strongly 
anisotropic and dependent on both mesons momenta and   polarizations.

Besides that,  weak magnetic field contributions for the anomalous couplings
were also presented with their momentum dependencies 
 in  the form factors, $F_{vja}^B(K,Q)$.
These 
corrections to the  anomalous vector mesons couplings to the axial currents
are 
reasonably
similar to the zero magnetic field case, $F_{vja}$.
These corrections are, however,
 numerically   suppressed by  the factor $eB/{M^*}^2$ that was assumed to be small.
Two different mechanisms that generate weak magnetic field  contributions 
for hadron couplings 
were considered.
Firstly he leading correction to the quark propagator - eq. (\ref{quark-prop})  -
and secondly the background photon overall  coupling to the quark-vector meson vertex.
Both mechanisms provide similar contributions that add to each other in 
eqs. (\ref{L-VJA-B}) and (\ref{mix-B}).
Vector-axial mesons mixing induced by the magnetic field was 
also found.

By following the same method considered  by Witten \cite{witten-1983}
for  Wess-Zumino terms,
conditions for the
 quantization of the anomalous couplings $F_{vja}$ were found.
The total axial current, however,  is only partially  conserved because of  Lagrangian quark masses
and of the ABJ anomaly. The vector current is not 
conserved due to the quark mass non degeneracy.
So,  it is not clear how to cope these small  symmetry breakings, dictated by 
well known 
low energy theorems, with
the topologically conserved terms in the action given by eqs. 
(\ref{quant}) or (\ref{quantiz-vja-V}).
They correspond to quantization conditions 
 that contain  
 the axial or vector  currents, or a components of them.
One way that, maybe, solve this
apparent contradiction would be to consider that the light vector or axial mesons
select components of, respectively, the axial or vector constituent quark currents that remain 
(topologically) preserved. 
This is not  investigated further in the present work.
For degenerate quark masses
the quantization 
condition for 
the  flavor singlet  axial meson coupling to the vector current, with $m_s$, is the same as the
flavor  triplet one, with $m_t$.
The  non degeneracy of quark masses  does not prevent the
quantization conditions to emerge,  such as  (\ref{quant}),  (\ref{quantiz-vja-V})  or 
(\ref{quantiz-vja-V-s}),
 it slightly changes its shape.
This case of  non degenerate quark mass will be addressed in another work.

\section*{Acknowledgments}

F.L.B. thanks  short discussions with G.I. Krein, C.D. Roberts and F.S. Navarra. 
The author  is member of
INCT-FNA,  Proc. 464898/2014-5,
and  he acknowledges partial support from 
CNPq-312072/2018-0 
and  
CNPq-421480/2018-1.


\begin{thebibliography}{00}


\bibitem{exp-r2A}
 K.L. Miller et al.,
{ Study of the reaction $\nu_\mu d \to \mu^-  p p_s$},
 Phys. Rev. D 26, 537 (1982); 
T. Kitagaki et al.,
{High-energy quasielastic $\nu_\mu n \to \mu^- p$ 
scattering in deuterium},
 Phys. Rev. D 28, 436 (1983).


\bibitem{gaillard-savage} 
J-M. Gaillard, G. Sauvage, {\it Hyperon Beta Decays},
Ann. Rev. Nucl. Part. Sci. {\bf 34}, 351 (1984).



\bibitem{choi-etal-93}
S. Choi, {\it et al},
{Axial and Pseudoscalar Nucleon 
Form Factors from Low Energy Pion Electroproduction},
 Phys. Rev. Lett. {\bf 71}, 3927 (1993).




\bibitem{bardin-etal-1981}
G. Bardin {\it et al } Measurement of the ortho para
 transition rate in the p mu p molecule and 
deduction of the pseudoscalar coupling constant $g_p^\mu$, 
Phys. Lett. {\bf 104 B},  320  (1981).



\bibitem{andreev-etal-2007}
V.A. Andreev, {\it et al}., (MuCap Collaboration),
{
Measurement of the Muon Capture 
Rate in Hydrogen Gas and Determination 
of the Proton’s Pseudoscalar Coupling $g_P$},
 Phys. Rev. Lett. {\bf 99}, 032002 (2007)




\bibitem{beise}
E.J.  Beise,{The Axial Form Factor of the Nucleon},
 Eur.Phys.J. A24S2, 43 (2005).



\bibitem{drechsel-walcher} D. Drechsel, Th. Walcher,
{Hadron structure at low Q$^2$},
Rev.Mod.Phys. 80, 731 (2008).


\bibitem{exp-bernard+E+meissner}
V. Bernard,  L. Elouadrihiri,
Ulf-G. Meissner,  {Axial structure of the nucleon},
J. Phys. G: Nucl. Part. Phys. {\bf 28}, R1  (2002).





\bibitem{maris-craig-tandy}
P. Maris, C. D. Roberts,
{ Dyson-Schwinger equations: A Tool for hadron physics},
Int. J. Mod. Phys. {\bf E 12},   297 (2003).
P. Tandy, 
{
Hadron physics from the global color model of QCD},
 Prog.Part.Nucl.Phys. {\bf 39}, 117 (1997).


\bibitem{eichmann-fischer}
G. Eichmann, C. S. Fischer, 
{Nucleon axial and pseudoscalar form factors from the covariant Faddeev equation},
Eur. Phys. J. {\bf A 48},  9 (2012).






\bibitem{ball-chiu}
J.S. Ball, T.W. Chiu, 
Analytic properties of the vertex function in gauge theories. I
Phys. Rev. D22, 2542  (1980).



\bibitem{segovia-etal}
Chen Chen, C. S.Fischer, C. D.Roberts, J. Segovia
Form factors of the nucleon axial current,
Phys. Lett. B815, 136150 (2021).




\bibitem{hlroberts-etal}
H.L.L. Roberts, et al,
$\pi$ and $\rho$ mesons, 
and their diquark partners, from a contact interaction,
Phys. Rev. C83, 065206 (2011).




\bibitem{vecmesff1} M.E. Carrillo-Serrano, W. Bentz, I. Cloet, A.W. Thomas,
$\rho$ meson form factors in a confining Nambu-Jona-Lasinio model,
Phys. Rev. C92, 015212 (2015).
E.P. Biernat, W. Schweiger, 
Electromagnetic $\rho$-meson form factors in point-form relativistic quantum mechanics,
Phys. Rev. C89, 055205 (2014).

\bibitem{vecmesff2}
V.Yu. Haurysh, V.V. Andreev, 
$\rho$-Meson Form-factors in Point form of Poincar\'e-Invariant
Quantum Mechanics,
Few-Body Syst. 62, 29 (2021).
J.P.B.C. de Melo, T. Frederico,
Covariant and light-front approaches to the $\rho$-meson electromagnetic form factors,
Phys. Rev. C55, 2043 (1997)



\bibitem{PRD-2018a}
F.L. Braghin, { Light vector and axial mesons effective couplings to constituent quarks},
Phys. Rev.  D 97, 054025 (2018);
Phys. Rev. D 101, 039901(E) (2020).


\bibitem{eLSM}
D. Parganlija, P. Kovacs, G. Wolf, F. Giacosa, D.H. Rischke,
Meson vacuum phenomenology in a three-flavor linear sigma
model with (axial-)vector mesons. Phys. Rev. D 87(1), 014011
(2013).
 


\bibitem{PDG}
P.A. Zyla et al. (Particle Data Group), Prog. Theor. Exp. Phys.   083C, 01 (2020).
K. Nakamura et al (Particle Data Group), 
 

\bibitem{kanchen-etal}
K. Chen, C.-Q. Pang,  X. Liu, T. Matsuki,
Light axial vector mesons,
Phys. Rev. D91, 074025 (2015).

\bibitem{du-zhao}
M.-C. Du, Q. Zhao, Comprehensive study of light axial vector mesons
with the presence of triangle singularity,
Phys. Rev. D104, 036008 (2021).





\bibitem{taudecay1}
D. Gomez Dumm, P. Roig, A. Pich, J. Portoles,
Hadron structure in $\tau \to KK\pi\nu_\tau$ decays,
Phys. Rev. D81, 034031 (2010).

\bibitem{taudecay2}
D. Gomez Dumm, P. Roig, A. Pich, J. Portoles,
$\tau \to \pi\pi\pi\nu_\tau$ 
decays and the a1(1260) off-shell width revisited
Physics Letters B 685, 158 (2010) .



\bibitem{wagner-leupold}
M. Wagner, S. Leupold,
Information on the structure of the $a_1$ from $\tau$-decay,
Phys. Rev. D78, 053001 (2018).


\bibitem{Mikhasenko-etal}
M. Mikhasenko et al,
Pole position of the A1(1260) from tau decay,
Phys. Rev. D98, 096021 (2018).

\bibitem{g2-problem} 
L. Cappiello, et al,
Axial-vector and pseudoscalar mesons in the hadronic light-by-light
contribution to the muon $(g-2)$,
Phys. Rev. D102, 016009 (2020).
and references therein.

\bibitem{f1285prod} 
P. Lebiedowicz, Otto Nachtmann, P. Salabura, A. Szczurek,
Exclusive $f_1(1285)$ meson production for energy ranges available at the
GSI-FAIR with HADES and PANDA,
Phys. Rev. D104, 034031 (2021).


\bibitem{sasaki-plb}
C. Sasaki, 
Signatures of chiral symmetry restoration in dilepton production,
Phys. Lett. B 801, 135172 (2020).


\bibitem{rho-a1-1}
E. Marco, R. Hofmann, W. Weise,
Note on finite temperature sum rules for vector and axial-vector
spectral functions,
Phys. Lett. B 530, 88 (2002).
M. Urban, M. Buballa, 
J. Wambach,
Temperature Dependence of $\rho$ and $a_1$-Meson Masses and Mixing
of Vector and Axial-Vector Correlators,
Phys. Rev. Lett. 88,  042002 (2002).




\bibitem{harada-etal}
M. Harada, C. Sasaki, W. Weise,
Vector–axial-vector mixing from a chiral effective field theory at finite temperature,
Phys. Rev. D78, 114003 (2008).
M. Harada, C. Sasaki,
Novel spectral broadening from vector–axial-vector mixing in dense matter
Phys. Rev. C80, 054912 (2009).



\bibitem{sasaki-etal1}
M. Harada, C. Sasaki, 
Effect of vector–axial-vector mixing to dilepton spectrum
in hot and/or dense matter,
arXiv:1003.0331v1 [hep-ph].



\bibitem{tripolt-etal}
R.A. Tripolt, et al,
Vector and axial-vector mesons in nuclear matter,
Phys. Rev. D104, 054005 (2021).

\bibitem{kovacs-etal}
Peter Kovacs, Zsolt Szep, Gyorgy Wolf,
Existence of the critical endpoint in the vector meson extended linear sigma model
Phys. Rev. D 93, 114014 (2016).




\bibitem{nuclear}
For example in: G. Chanfray, J. Magueron, 
Contribution of the $\rho$ 
meson and quark substructure to the nuclear spin-orbit potential
Phys. Rev. C102, 024331 (2020).



\bibitem{adler-bardeen}
S.L. Adler, W.A. Bardeen, Absence of Higher-order Corrections
in the Anomalous Axial-Vector Divergence equation,
Phys. Rev. 182, 1517 (1969).
W.A. Bardeen, Anomalous Ward Identities in Spinor Field Theories,
Phys. Rev. 184, 1848 (1969).

\bibitem{witten-1983}
E. Witten, Global aspects of Current Algebra, 
Nucl. Phys. B223, 422 (1983).



\bibitem{gomm-etal}
H. Gomm,  O. Kaymakcalan, J. Schechter, 
Anomalous spin-1-meson decays from the gauged Wess-Zumino term,
Phys. Rev. D30, 2345 (1984).


\bibitem{vecmes}
U.-G. Meissner,
Low-energy hadron physics from 
effective chiral Lagrangians,
Phys. Rept. 161, 213 (1988).

\bibitem{hohler-rapp}
P.M. Hohler, R. Rapp,
Realistic implementation of massive Yang-Mills theory for $\rho$
 and $a_1$ mesons,
Phys. Rev. D89, 125013 (2014).



\bibitem{ER-1986}
D. Ebert, H. Reinhardt,
Effective chiral hadron lagrangian with 
anomalies and skyrme terms from quark flavour dynamics,
Nucl. Phys. B 271,  188 (1986).




 \bibitem{sigmamodel}
P. Ko, S. Rudaz, 
Phenomenology of scalar and vector mesons in the linear $\sigma$ model,
Phys. Rev. D50, (1994).



\bibitem{vecmes-photo}
V. Mathieu, et al, Vector meson photoproduction with a linearly polarized beam,
Phys. Rev D97, 094003 (2018).

\bibitem{roy-thesis}
P. Roy, PhD dissertation,
Measurement of polarization observables in vector meson photonproduction
using a transversely polarized frozen-spin target at CLAS, JLAB, 
Florida State University,
(2016).


\bibitem{lebiedowicz-etal} 
P. Lebiedowicz, et al,
Central exclusive diffractive production of axial-vector 
$f_1(1285)$ and $f_1(1420)$
 mesons in proton-proton collisions,
Phys. Rev. D102, 114003 (2020).




\bibitem{ECQM-miller-etal}
T. J. Hobbs, M. Alberg, G. A. Miller,
A Euclidean bridge to the relativistic constituent quark model,
Phys. Rev. C 95, 035205 (2017)



\bibitem{diquarks-craig}
Chen Chen {\it et al},
Nucleon axial-vector and pseudoscalar form factors, and PCAC relations,
arXiv:2103.02054

\bibitem{5quarks}
J.-B. Wang, The axial charges of proton within an extended chiral
constituent quark model, 
arXiv:2106.00866v1 [hep-ph].


\bibitem{PRC1}
C.D. Roberts, R.T. Cahill, J. Praschifka, 
The effective action for the Goldstone modes in a
global colour symmetry model of QCD,
 Ann. of Phys. 188,   20 (1988).




\bibitem{tandy}
P.C. Tandy, Hadron Physics from the Global Color Model of QCD,
Prog.Part.Nucl.Phys. 39, 117 (1997).



\bibitem{klevansky}
S. P. Klevansky 
The Nambu-Jona-Lasinio model of quantum chromodynamics,
 Rev. Mod. Phys. 64,    649 (1992).


\bibitem{weise-vogl}
U. Vogl, W. Weise, 
{The Nambu and Jona-Lasinio model: Its implications for Hadrons and Nuclei},
 Progr. in Part. and Nucl. Phys. 27, 195 (1991).




\bibitem{BRST-2010}
S.J. Brodsky, C.D. Roberts, R. Schrock, P.C. Tandy,
New perspectives on the quark condensate,
Phys. Rev. C82, 022201 (2010).

\bibitem{weinberg-axial}
S. Weinberg, Why do quarks behave like bare Dirac particles?,
Phys. Rev. Lett. 65, 1181 (1990).
S. Weinberg, 
Axial vector coupling of the quark, 
Phys. Rev. Lett. 67, 3473 (1991).



\bibitem{weinberg-2010}
S. Weinberg,   
{ Pions in Large 
N
 Quantum Chromodynamics},
Phys. Rev. Lett.  105,    261601 (2010).   



\bibitem{PRD-2019}
F.L. Braghin, 
Pion Constituent Quark Couplings strong form factors:
A dynamical approach, 
Phys. Rev. D 99,   014001 (2019).


\bibitem{EPJA-2016}
 F.L. Braghin,  
Quark and pion effective couplings from polarization effects,
 Eur. Phys. Journ.  A 52,   134 (2016).


\bibitem{EPJA-2018}
F.L. Braghin, 
Low energy constituent quark and pion effective couplings in
a weak external magnetic field,
Eur. Phys. J. A54,  45 (2018).

\bibitem{review-B-1}
V.A. Miransky, I.A. Shovkovy,
Quantum field theory in a magnetic field: from quantum
chromodynamics to graphene and Dirac semimetals, Phys. Rep. 576, 1 (2015)

\bibitem{review-B-2}
J.O. Andersen, W.R. Naylor. A. Tranberg,
 Phase diagram of QCD in a magnetic field: a
review, Rev. Mod. Phys. 88, 025001  (2016).



\bibitem{PRD-2018b}
F.L. Braghin,
{ Constituent quark-light vector mesons effective couplings
in a weak background magnetic field},
 Phys. Rev. D 97,  014022 (2018);
Phys. Rev. D 101, 039902(E) (2020)


 
\bibitem{JPG-2020a}
W.F. de Sousa, F.L.Braghin, 
 Form factors for  pions couplings  to  constituent quarks
 under  weak magnetic field,
Journ. of Phys.  G 47, 045110 (2020).



\bibitem{JPG-2020b}
F.L. Braghin,
Weak magnetic field corrections to light
vector or axial mesons mixings and vector
meson dominance,
J. Phys. G: Nucl. Part. Phys. 47,   115102  (2020).



\bibitem{SD-rainbow}
D. Binosi, L. Chang, J. Papavassiliou, C.D. Roberts, 
{
Bridging a gap between continuum-QCD and ab initio predictions of hadron observables},
Phys. Lett.   B 742,  183  (2015) and references therein.


\bibitem{kondo}
K.-I. Kondo,
{Abelian-projected effective gauge theory of QCD 
with asymptotic freedom and quark confinement}
 Phys. Rev.  D 57, 7467 (1998) .


\bibitem{cornwall} J. M. Cornwall,
{Entropy, confinement, and chiral symmetry breaking},
Phys. Rev.  D 83, 076001 (2011).

\bibitem{dcsb+traceanomaly} 
 X.-D. Ji,
QCD Analysis of the Mass Structure of the Nucleon,
 Phys. Rev. Lett. 74, 1071 (1995),
arXiv:hep-ph/9410274.
W. Kou, R. Wang, X. Chen,
Trace Anomaly of Proton Mass with Vector Meson Near-Thresholds Photoproduction
Data,.
arXiv:2103.10017v2 [hep-ph].
The ideas of EHM: for example in
C.D. Roberts, S.M. Schmidt,  Reflections upon the emergence of hadronic mass,
Eur. Phys. J. Special Topics 229, 3319  (2020).


\bibitem{lowdon}
P. Lowdon, 
Nonperturbative structure 
of the photon and gluon propagators,
Phys. Rev. D96, 065013 (2017).


\bibitem{FLB-2021a}
F.L. Braghin,  Flavor-dependent U(3) Nambu–Jona-Lasinio coupling constant,
  Phys. Rev. D103, 094028 (2021).


\bibitem{vec-ax1}
D. Parganlija, P. Kovacs, G. Wolf, F.
Giacosa, and D. H. Rischke, 
Meson vacuum phenomenology in a three-flavor linear sigma model with (axial-)vector mesons,
Phys. Rev. D 87, 014011
(2013).



\bibitem{PRD-2014}
A. Paulo Jr., F.L. Braghin,
Vacuum polarization corrections to low energy quark effective couplings,
Phys. Rev. D90, 014049 (2014).


\bibitem{integr-pathint}
 M. Gaillard, The Effective One Loop Lagrangian With Derivative Couplings, Nucl. Phys. B 268,
669 
(1986).
 L.-H. Chan, Derivative Expansion for the One Loop Effective Actions With Internal Symmetry,
Phys. Rev. Lett. 57, 1199 (1986).
 O. Cheyette, Effective Action for the Standard Model With Large Higgs Mass, Nucl. Phys. B 297,
183
(1988).

\bibitem{vec-axial-mesons}
C. A. Ballon Bayona {\it et al},
Form factors of vector and axial-vector mesons in holographic D4-D8 model,
Journal of High Energy Physics  2010,  52 (2010).


\bibitem{unal-etal}
Y. Unal, A. Ku\c cukarslan, S. Scherer,
Contribution of the a1 meson to the axial nucleon-to-delta
transition form factors,
Phys. Rev. 90, 014012 (2019).


\bibitem{ecker-etal-1989}
G. Ecker, J. Gasser, J. Leutwyler, A. Pich, E. de Rafael,
Chiral Lagrangians for massive spin 1 fields,
Phys. Lett. B223, 425 (1989).


\bibitem{roca-etal-2004}
L. Roca, J.E. Palomar, E. Oset, 
Decay of axial-vector mesons into VP and P$\gamma$,
Phys. Rev. D 70, 094006 (2004).


\bibitem{birse} 
M.C. Birse,
Effective chiral Lagrangians for spin-1 mesons, 
Z. Phys. A 355, 231–246 (1996).


\bibitem{tandy-maris}
P. Maris, P.C. Tandy,
{\it Bethe-Salpeter study of vector meson masses and decay constants},
 Phys. Rev. C 60, 055214 (1999).


\bibitem{curvature-mass-kovacs-etal}
G. Kov\'acs, P. Kov\'acs, Z. Sz\'ep
One-loop constituent quark contributions to the vector
and axial-vector meson curvature mass
Phys. Rev. D104, 056013 (2021).



\bibitem{kim-lee-symmetric-vecmesmass}
J. Kim, S.H. Lee,
Vector meson mass in the chiral symmetry restored vacuum,
Phys. Rev. D103, 
 103, L051501 (2021)


\bibitem{kaiser-etal}
N. Kaiser, U.-G. Meissner,
Generalized hidden symmetry for low-energy hadron physics,
Nucl. Phys. A519, 671 (1990).




\bibitem{skyrme}
I. Zahed, G.E. Brown, 
The Skyrme model,
Phys. Rept.  142, 1 (1986).
Balachandran et al,
A.P. Balachandran, V.P. Nair, S.C. Rajeev and A. Stern, 
Phys. Rev. Lett. 49 (1982) 1124; Phys. Rev. D27 (1983) 1153



\bibitem{weinberg-book}
S. Weinberg, The Quantum Theory of Fields Vol. II 
Cambridge
University Press, Cambridge,  (1996).


\bibitem{navarra+est}
 I. Danhoni, F. S. Navarra,
Magnetic field in relativistic heavy ion collisions: testing the classical approximation,
Phys. Rev. C 103, 024902 (2021).


\bibitem{navarra+est2}
I. Danhoni, F. S. Navarra,
Magnetic excitation in relativistic heavy ion collisions,
Phys. Lett. B 805, 135463  (2020).


\bibitem{weak-B}
 T.-K.  Chyi  et al,
 Weak-field expansion for processes in a homogeneous background magnetic
field, Phys. Rev. D 62 105014 (2000).


\bibitem{igor1}
E.V. Gorbar, V.A. Miransky, I.A. Shovkovy,  X. Wang,
 Radiative corrections to chiral
separation effect in QED,
 Phys. Rev. D 88 025025 (2013).




\bibitem{ooliveira-etal}
P. Costa, O. Oliveira, P. J. Silva, What does low energy
physics tell us about the zero momentum gluon propagator,
Phys. Lett. B 695, 454 (2011).


\bibitem{nucleon-eff}
See for example: S. Pacetti, R. B. Ferroli, E. Tomasi-Gustafsson,
Proton electromagnetic form factors: Basic notions, present
achievements and future perspectives,
Phys. Rep. 550-551, 1 (2015),
and references therein.



\bibitem{ballon-etal}
A. Ballon-Bayona, G. Krein, and C. Miller, Phys. Rev. D 96,
014017 (2017).

\bibitem{krutov-etal}
A. F. Krutov, R. G. Polezhaev, and V. E. Troitsky, Phys. Rev.
D 93, 036007 (2016) and references therein.

\bibitem{bhagwat-maris} M.S. Bhagwat, P. Maris, 
Vector meson form factors and their quark-mass dependence,
Phys. Rev. C77, 025203 (2008).









 \end{thebibliography}
\end{document}